\RequirePackage{rotating}
\documentclass[a4paper,fleqn]{cas-dc}

\usepackage{xcolor}
\usepackage{caption}
\usepackage{float}
\usepackage{amsmath}

\usepackage{bm}
\usepackage{multirow}
\usepackage{url,hyperref,lineno,microtype,subcaption}
\usepackage[onehalfspacing]{setspace}
\usepackage{textgreek}
\usepackage{adjustbox}
\usepackage{tabularx}
\usepackage{lipsum}
\usepackage{soul}
\usepackage{multicol}
\usepackage{placeins}
\usepackage{afterpage}
\usepackage{rotating}
\usepackage[numbers, sort&compress]{natbib}
\usepackage{nomencl}
\makenomenclature
\usepackage[acronym]{glossaries}
\makeglossaries
\usepackage{framed} 

\newcommand{\rev}[1]{\textcolor{black}{#1}}
\newcommand{\revb}[1]{\textcolor{black}{#1}}
\newcommand{\revaa}[1]{\textcolor{black}{#1}}
\newcommand{\revaasecond}[1]{\textcolor{black}{#1}}
\newcommand\blfootnote[1]{%
  \begingroup
  \renewcommand\thefootnote{}\footnote{#1}%
  \addtocounter{footnote}{-1}%
  \endgroup
}

\def\tsc#1{\csdef{#1}{\textsc{\lowercase{#1}}\xspace}}
\tsc{WGM}
\tsc{QE}
\tsc{EP}
\tsc{PMS}
\tsc{BEC}
\tsc{DE}

\begin{document}
\let\WriteBookmarks\relax
\def\floatpagepagefraction{1}
\def\textpagefraction{.001}

\shorttitle{Hybrid lunar ISRU plant: a comparative analysis with carbothermal reduction and water extraction}

\shortauthors{K. Ikeya et~al.}

\title [mode = title]{Hybrid lunar ISRU plant: a comparative analysis with carbothermal reduction and water extraction}                      



%
\author[1]{Kosuke Ikeya}[type=editor,
                        orcid=0000-0003-1172-1527]

\cormark[1]

\fnmark[1]

\ead{k.ikeya22@imperial.ac.uk}

\ead[url]{https://profiles.imperial.ac.uk/k.ikeya22}

\credit{Conceptualization, Methodology, Software, Data curation, Formal analysis, Investigation, Writing - Original draft preparation, Writing – review \& editing}

\affiliation[1]{organization={Department of Earth Science and Engineering, Imperial College London},
    addressline={Prince Consort Road}, 
    city={London},
    postcode={SW7 2AZ}, 
    country={The United Kingdom}}





\author[2]{Francisco J. Guerrero-Gonzalez}[]

\affiliation[2]{organization={Professorship of Lunar and Planetary Exploration, Technical University of Munich},
    addressline={Lise-Meitner-Str. 9}, 
    city={Ottobrunn},
    postcode={85521}, 
    country={Germany}}
    
\credit{Conceptualization, Methodology, Software, Data curation, Writing - Original draft preparation, Writing – review \& editing}

\author[3]{Luca Kiewiet}[]
\affiliation[3]{organization={Institute of Space Systems, German Aerospace Center},
    addressline={Robert-Hooke-Straße 7}, 
    city={Bremen},
    postcode={28359}, 
    country={Germany}}
    
\credit{Conceptualization, Methodology, Software, Writing - Original draft preparation}



\author[4]{Michel-Alexandre Cardin}
\credit{Conceptualizationn, Writing – review \& editing, Supervision}
\author[1]{Jan Cilliers}
\credit{Conceptualization, Supervision, Funding acquisition}
\author[1]{Stanley Starr}
\credit{Conceptualization, Writing – review \& editing, Supervision}

\affiliation[4]{organization={Dyson School of Design Engineering, Imperial College London},
            addressline={Imperial College Road}, 
            city={London},
            postcode={SW7 2DB}, 
            country={United Kingdom}}

\author[5]{Kathryn Hadler}
\affiliation[5]{organization={European Space Resources Innovation Centre (ESRIC), Luxembourg Institute of Science and Technology (LIST)},
            addressline={5 Avenue des Hauts-Fourneaux}, 
            postcode={L-4362}, 
            country={Luxembourg}}
\credit{Conceptualization, Supervision}



\cortext[cor1]{Corresponding author}



\begin{abstract}
To establish a self-sustained human presence in space and to explore deeper into the solar system, extensive research has been conducted on In-Situ Resource Utilization (ISRU) systems. Past studies have proposed and researched many technologies to produce oxygen from regolith, such as carbothermal reduction and water extraction from icy regolith, to utilize it for astronauts' life support and as the propellant of space systems. However, determining the most promising technology remains challenging due to uncertainties in the lunar environment and processing methods. To better understand the lunar environment and ISRU operations, it is crucial to gather more information. Motivated by this need for information gathering, this paper proposes a new ISRU plant architecture integrating carbothermal reduction of dry regolith and water extraction from icy regolith. 
Two different hybrid plant architectures integrating both technologies (1) in parallel and (2) in series are examined. The former involves mining and processing in both a Permanently Shadowed Region (PSR) and a peak of eternal light in parallel, while the latter solely mines in a PSR. In this series hybrid architecture, the dry regolith tailings from water extraction are further processed by carbothermal reduction.
This paper conducts a comparative analysis of the landed mass and required power of each plant architecture utilizing subsystem-level models. Furthermore, based on uncertain parameters such as resource content in regolith, the potential performance range of each plant was discovered through Monte Carlo simulations. The result indicates the benefit of the series hybrid architecture in terms of regolith excavation rate \revaasecond{and power consumption}, while its mass cost seems the highest among the studied architectures.
\end{abstract}



\begin{keywords}
in-situ resource utilization  \sep lunar regolith \sep oxygen extraction  \sep carbothermal reduction \sep icy regolith  \sep uncertainty analysis
\end{keywords}

\maketitle

\blfootnote{\revaa{Abbreviations: CHX, condensing
heat exchanger; ConOps, concept of operations; COTS, commercial-
off-the-shelf; CR, carbothermal reduction; ISRU; in-situ resource utilization; PEL, peak of eternal light; PEM, proton
exchange membrane; PH, parallel hybrid; PSR, permanently shadowed region; LH\textsubscript{2}, liquid hydrogen; LOX, liquid oxygen; SH, series hybrid; SOXE, solid oxide electrolysis; WE, water extraction}}








\section{Introduction}
In-Situ Resource Utilization (ISRU), also known as \revaa{space resource utilization}, aims to process and utilize local extraterrestrial resources to enable long-term human presence beyond low Earth orbit. Resources on our nearest planetary bodies, i.e., the Moon and Mars, include water in the form of ice or hydrated minerals \citep{Crawford.2015, Starr.2020, Anand.2012}, atmospheric carbon dioxide \citep{Starr.2020}, and regolith, which is a source of oxygen and metals \citep{Crawford.2015, Anand.2012}. These resources can be processed by a variety of thermo- and electrochemical techniques \citep{Starr.2020, Schluter.2020} to produce life support consumables (e.g., H\textsubscript{2}O, O\textsubscript{2}), rocket fuel propellant (e.g.,  CH\textsubscript{4}, H\textsubscript{2}, O\textsubscript{2}), or infrastructure and spare parts (e.g., regolith, metals). Among various ISRU activities, lunar ISRU has been increasingly gathering a large amount of attention in response to the concurrent Artemis program and other initiatives of going back to the Moon.

For lunar ISRU, several architectures for the production of these propellant and life-support consumables have been considered: (1) oxygen extraction from dry-regolith processing with reducing agents, such as hydrogen~\cite{Sargeant.2021,Lee.2013} and methane~\cite{Linne.2021, Rice.1996, Troisi.2022, White.2023, Prinetto.2023}, \revb{(2) oxygen extraction via electrolysis of molten regolith \cite{Sibille2009} and in molten salt \cite{Fray2002, Ono2002},} and (3) hydrogen and oxygen extraction from lunar icy regolith~\cite{Sowers.2019, Pelech.2019,Purrington.2022, Schieber.2022,He.2021, Liu.2023,Collins.2023,Kiewiet.2022, Kleinhenz.2020, Cole.2023}.
Choosing the most suitable architecture is crucial for successful and economically viable lunar ISRU \citep{Anand.2012, Schluter.2020, Taylor.1992b}. However,
it is highly unpredictable if either architecture would perform significantly better than the other due to uncertainties in the resource and operation.
Building an ISRU plant dedicated to only one of these technologies might be suboptimal when these inherent uncertainties unfold.

Recognizing uncertainty, this paper proposes a novel architecture: a hybrid ISRU plant, where liquid oxygen (LOX) and liquid hydrogen (LH\textsubscript{2}) are coproduced from both dry and icy regolith on the Moon.
\revb{Hybrid architectures that utilize multiple resources simultaneously have previously been proposed only for Mars ISRU~\cite{Kleinhenz.2017, Chen.2020}; the benefits and costs of processing both dry and icy regolith on the Moon have not been thoroughly explored.}

Two different types of hybrid lunar ISRU architectures are proposed. In the first architecture, \revaa{Carbothermal Reduction (CR)} and direct \revaa{Water Extraction (WE)} from icy regolith occur simultaneously in parallel. In the other architecture, these two processes are integrated in series, i.e., the remaining dry regolith after \revaa{WE} is further processed through \revaa{CR}. The Concept of Operations (ConOps) suitable for each architecture is carefully defined.
Although these proposed hybrid architectures add complexity to the overall system, they can mitigate the risk of failure by avoiding reliance on a single extraction technology.

This paper further compares the mass and power budgets, along with other system performance indicators, of the proposed hybrid plant architecture with single technology architectures: \revaa{CR} of dry regolith and direct \revaa{WE} from icy regolith. 
A holistic and multidisciplinary approach that models end-to-end ISRU production plants 
is utilized to understand the benefits and drawbacks of specific technologies and processing techniques \citep{Hadler.2020}.

\revb{Although there have been an increasing number of missions to the Moon and studies about it~\cite{Gaddis2023}, the knowledge available about the lunar environment still contains large uncertainty \cite{Cilliers.2020}. The effects of this uncertainty on the performance of a plant and its operation 
have not been discussed enough \cite{Cilliers2023}.}
\revb{Therefore,} to examine the performance of these different plant architectures in various conditions, Monte Carlo simulations explicitly address resource and operational uncertainty.
The results highlight the benefits and drawbacks inherent to each plant design.

The rest of the paper is organized as follows: 
Section~\ref{sec:literature} summarizes past studies in hybrid ISRU architectures, mass and power estimation of ISRU systems, and uncertainty related to lunar ISRU highlighting the research gap addressed by this paper.
Section~\ref{sec:study_approach} introduces the hybrid ISRU production plant concept, including assumptions made and ConOps. 
In Section~\ref{sec:isru_model}, the modeling approach of this work for the mass and power estimations is explained. The results of the mass and power estimations are discussed further in Section~\ref{sec:result_deterministic}.
Section~\ref{sec:uncertainty_in_lunar_isru} introduces the intrinsic uncertainty of the lunar environment and ISRU systems, and its effects on the performance of each ISRU plant operation. 
Section~\ref{sec:discussion} discusses the potential benefits and risks of each plant design as well as \revaa{limitations of this study and directions for future work}. Finally, concluding remarks are outlined in Section~\ref{sec:conclusion}.

\section{Background and related Work}
\label{sec:literature}
\subsection{Hybrid ISRU architecture}
Compared to the conventional single-technology ISRU architectures, hybrid architectures combining multiple process technologies have not been well-researched.
Kleinhenz and Paz~\cite{Kleinhenz.2017} proposed the idea of utilizing \rev{the Martian atmosphere and subsurface water ice}  simultaneously for Mars ISRU. \rev{This study revealed a potential reduction in landed mass compared to the conventional oxygen production architecture for Mars return vehicle propellant by producing both liquid methane and liquid oxygen.}

Chen et al.~\cite{Chen.2020} also showed the benefit \rev{of employing both hydrogen reduction and \revaa{WE} from Martian soil} through the optimization of technology selection in terms of crewed Mars mission cost. \rev{Even though this study showed the economic benefit, they also noted the potential increase in the complexity of its development, deployment, and operation.}

While it is likely that these hybrid architectures add significant complexity, these past studies indicate some potential benefits.
\revb{These drawbacks and advantages require more in-depth analyses to fully assess these architectures compared to other ISRU options.}
Lunar ISRU utilizing both dry and icy regolith, especially, needs more understanding given the increasing international interest in human lunar exploration. However, the performance of hybrid lunar ISRU is yet to be researched.

\subsection{Mass and power estimation for ISRU}
\label{subsec:lit_mass_power}
The estimation of mass and power budgets of all the necessary subsystems is essential before considering ISRU as a viable alternative to transporting all material resources from Earth \citep{GuerreroGonzalez.2023}. \rev{Therefore, the estimation has been conducted extensively through (mainly) computer modeling with different fidelity levels.}

\rev{As an example of relatively low-fidelity modeling, }Chen et al.~\cite{Chen.2020} developed a database of subsystem-level specific masses and powers \rev{based on literature, such as \revaa{Santiago-Maldonado and Linne}~\cite{Santiago-Maldonado.2007},} and integrated it into an Earth-Moon-Mars space resources logistics network. This approach analyzed the effect of ISRU deployment location (i.e., the Moon or a near-Earth object) regarding system mass and monetary mission cost.

\rev{While these models can capture the trend in mass and power for different architectures, they often overlook the detailed ConOps, potentially missing some key subsystems and components to be considered. Recently, more detailed models considering ConOps carefully have been developed.} Linne et al.~\cite{Linne.2021} modeled in detail the extraction of oxygen via \revaa{CR} of dry regolith to identify the feasibility of deploying such a plant considering the capacity of a lunar lander.
Guerrero-Gonzalez and Zabel~\cite{GuerreroGonzalez.2023} also accounted for metal processing by comparing three different ISRU techniques, hydrogen reduction, molten regolith electrolysis, and molten salt electrolysis, and estimating the mass and required power of each subsystem.
Kiewiet et al.~\cite{Kiewiet.2022} modeled \revaa{WE} from icy regolith via three different heating methods and conducted a trade-off study. Kleinhenz and Paz~\cite{Kleinhenz.2020} compared the system mass and power of two different architectures: the extraction of both H\textsubscript{2} and O\textsubscript{2} from water ice in the lunar southern polar region and solely O\textsubscript{2} based on the work developed by Linne et al.~\cite{Linne.2021}.

\subsection{Recognizing uncertainty in ISRU}
Although the studies \rev{in Section~\ref{subsec:lit_mass_power}} aimed to quantitatively analyze the most promising ISRU technologies, it should be noted that the comparisons made in them are based on deterministic assumptions for some key parameters such as resource or operational availability. However, as Cilliers et al.~\cite{Cilliers.2020} showed \rev{by considering uncertainty in many aspects, such as resource content}, such oversimplification by ignoring the uncertainty of key parameters might change the requirements for the ConOps and, therefore, overlooks the potential risk of underperformance.
Takubo et al.~\cite{Takubo.2022} considered the uncertainty in the water production rate and the yearly decay rate in this production. Although the work successfully integrated these parameters into a spaceflight campaign design, the uncertainty recognized in this work is limited. 
\rev{Malone et al.~\cite{Malone.2022} also recognized uncertainty in the entire mining process, such as potential subsystem technological malfunction and power generation. Their study integrated the recognized uncertainty into a Comprehensive Lunar Mining Simulator to help decision-makers through a serious game approach.}

\rev{None of these studies have compared different ISRU technologies considering the effect of the uncertain lunar environment and ISRU operations. Due to the large difference between dry and icy regolith processing, potential benefits and risks can be easily overlooked when ignoring the uncertainty. Inspired by \revaa{Cilliers et al.}~\cite{Cilliers.2020}, this paper models the distributions of uncertain parameters and examines their effects on plant performance indicators such as regolith excavation rate.}
\section{Hybrid lunar ISRU plant concept}\label{sec:study_approach}
\rev{This section first proposes two different hybrid lunar ISRU architectures \revaa{combining two extraction technologies}. Technologies selected for further analyses and the detailed ConOps are explained in Sec.~\ref{sec:technology_selection} and \ref{sec:concept_operations}, respectively.}
\subsection{Overview and assumptions}\label{sec:assumptions}
The lunar South Pole was selected as an ideal location to place the hybrid ISRU production plant due to the increasing evidence of volatile and water ice presence in its Permanently Shadowed Regions (PSRs) \cite{Colaprete.2010, Hayne.2015, Li.2018} and the existence of Peaks of Eternal Light (PELs) that provide extended periods of solar illumination \citep{Mazarico.2011, Zuniga.2019}.

\begin{figure*}[h!]
\centering
     \begin{subfigure}[b]{0.4\textwidth}
         \centering
         \includegraphics[width=\linewidth]{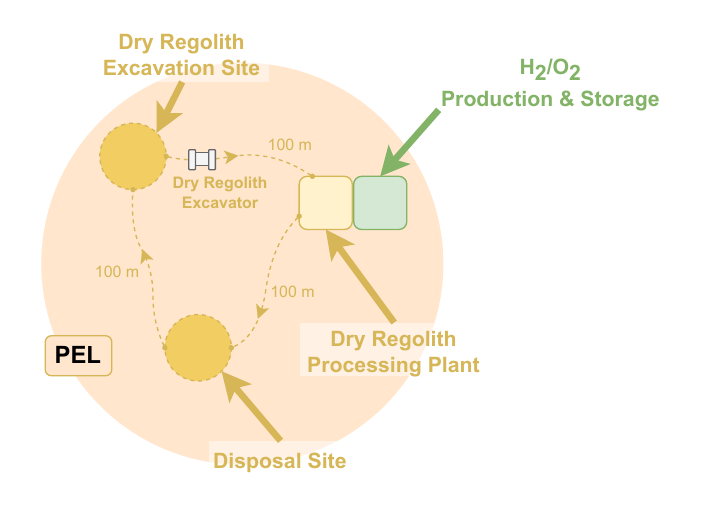}
         \caption{\revaa{Typical dry regolith processing architecture.}}
         \label{fig:cr}
     \end{subfigure}
     \begin{subfigure}[b]{0.52\textwidth}
         \centering
         \includegraphics[width=\linewidth]{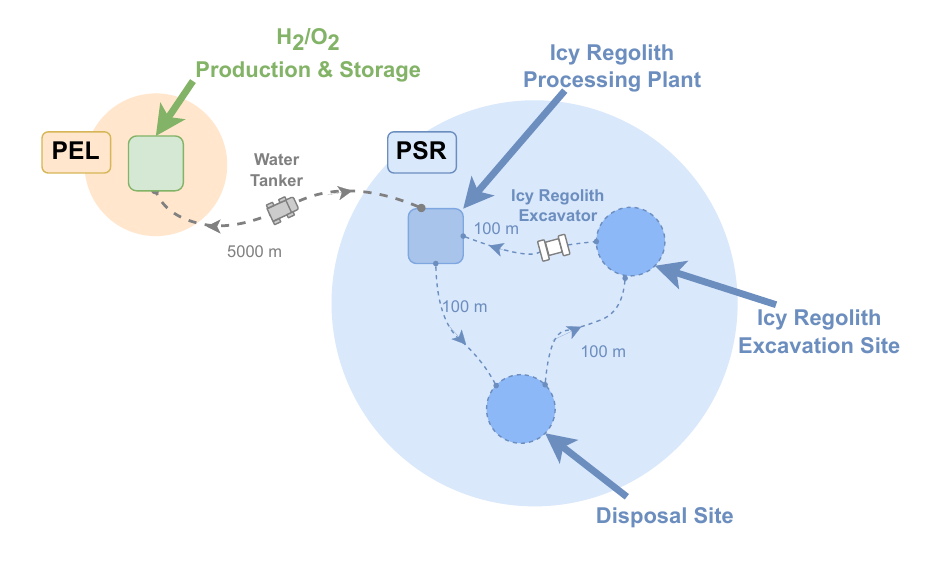}
         \caption{\revaa{Direct water extraction architecture based on Kleinhenz and Paz~\cite{Kleinhenz.2020}.}}
         \label{fig:we}
     \end{subfigure}
     \vfill
     \begin{subfigure}[b]{\textwidth}
         \centering
         \includegraphics[width=0.72\linewidth]{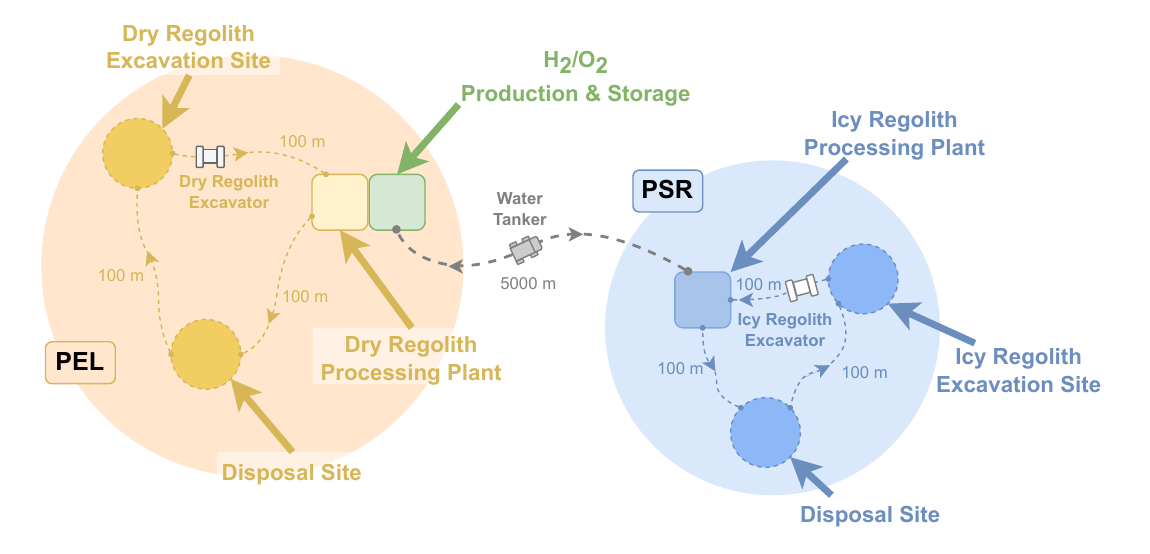}
         \caption{Parallel hybrid architecture. Dry regolith and icy regolith are processed simultaneously in different sites.}
         \label{fig:type_a}
     \end{subfigure}
     \vfill
     \begin{subfigure}[b]{\textwidth}
         \centering
         \includegraphics[width=0.52\linewidth]{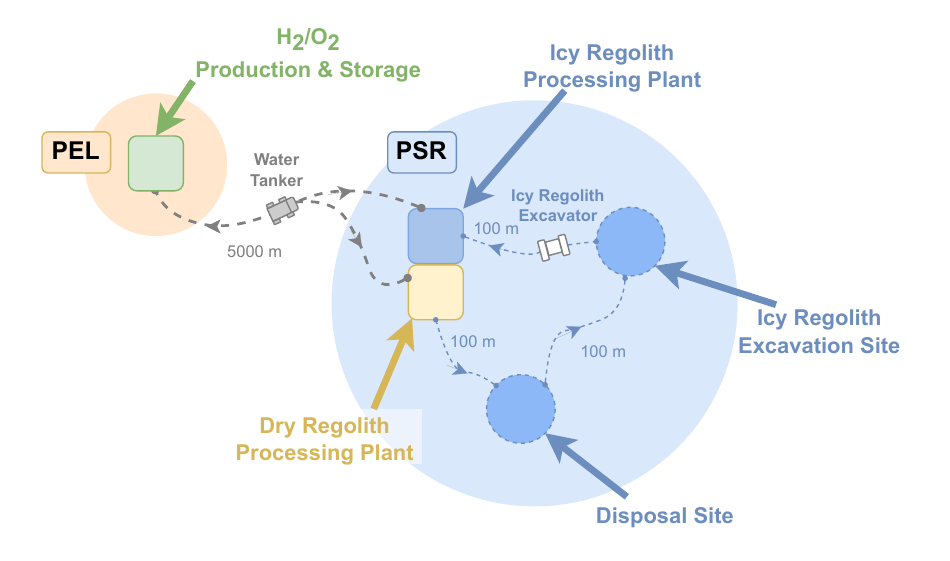}
         \caption{Series hybrid architecture. Dry regolith from \revaa{WE} process is further processed in a PSR.}
         \label{fig:type_b}
     \end{subfigure}
\caption{\revaa{ISRU production plant architectures considered in this paper.}}
\label{fig:CONOPS}
\end{figure*}

\revaa{Figure \ref{fig:CONOPS} depicts rather conventional single-technology ISRU architectures as well as hybrid architectures proposed in this paper. A typical dry regolith processing ISRU architecture, such as CR or hydrogen reduction, has a regolith excavation site, and a disposal site as well as a processing site (Fig.~\ref{fig:cr}). For WE from icy regolith, among various different kinds of architectures, an architecture suggested by Kleinhenz and Paz~\cite{Kleinhenz.2020} is depicted in Fig.~\ref{fig:we}. In this architecture, water tankers transport water from a PSR to a PEL to further conduct electrolysis and liquefaction. For other WE technologies, please refer to Section \ref{sec:technology_selection}.}

Two different types of hybrid production plants are considered in this paper. \revb{The first architecture considered is a }\revaa{Parallel Hybrid (PH)} plant that excavates and extracts resources from both PEL and PSR sites (Fig.~\ref{fig:type_a}). In the PEL, dry regolith is excavated and processed to extract oxygen. Simultaneously in the PSR, water is extracted from icy regolith. \revaa{Following the \revaa{WE} plant design proposed by Kleinhenz and Paz~\cite{Kleinhenz.2020},} the extracted water is transported from the PSR to the PEL by a mobile water transportation solution, where the water is then electrolyzed into O\textsubscript{2} and H\textsubscript{2}. Thermochemical extraction processes, such as carbothermal or hydrogen reduction of dry regolith, produce water as an intermediate product, and therefore, water electrolysis and storage of oxygen and hydrogen can be shared between dry and icy regolith processing.
 
\revb{The second architecture considered is a} \revaa{Series Hybrid (SH)}, \revb{where} dry regolith tailings from the \revaa{WE} of icy regolith are further processed directly in the PSR (Fig.~\ref{fig:type_b}).
Similar to the \revaa{WE and PH} architectures, the extracted and produced water is transported to a PEL for further processing. The detailed ConOps of both hybrid plants are further described in Section~\ref{sec:concept_operations}.

\revaasecond{Alternatively, transporting regolith excavated in a PSR, as well as water from WE, to a PEL for CR and further process is possible. This approach can reduce the number of excavators compared to the PH architecture. However, carrying excavated regolith from the PSR to the PEL requires a regolith transporting solution, such as a tanker. The results from Kleinhenz and Paz \cite{Kleinhenz.2020} indicate the mass of such tankers can be much more than the saved mass from excavators, and therefore this option is not considered further.}

Another possibility would be targeting micro cold traps for ice mining instead of accessing deeper permanently shadowed craters where larger water deposits might be found. However, the resource availability and extraction viability of these small PSRs are still highly unknown \citep{Hayne.2021}. Therefore, this scenario has not been considered part of the study.

Although identifying a specific site within the lunar South Pole was out of the scope of this work, previous studies \citep{Kleinhenz.2020, Kaschubek.2021, Sathyan.2024, Kleinhenz.2022} have identified traverse distances of around \revaa{0.65}-8.5 km between potential human landing system sites, permanently shadowed craters where acceptable water ice deposits might be found, and ridges with prolonged sunlight availability. A preliminary analysis demonstrated that mass and power budgets were not sensitive to changes in the traverse distance. Therefore, an average traverse distance of 5 km between PSRs and PELs is employed in this study as depicted in Fig. \ref{fig:CONOPS}.

As a baseline case scenario, the ISRU plant is assumed to produce 10 \revaa{t} of O\textsubscript{2} and 1.25 \revaa{t} of H\textsubscript{2} per year. This 8:1 ratio corresponds to the stoichiometric mass ratio of water to maximize the use of local resources for various purposes. It is worth noting that, for hydrolox lunar ascent vehicle propellant, the fuel mixture mass ratio is generally close to 6:1. Some past studies employed this 6:1 ratio or a similar ratio as their required products ratio allowing an excess of oxygen \cite{Kleinhenz.2020, Sanders.2010}.

The integrated hybrid production plant model in this paper includes all the necessary ISRU subsystems, from regolith excavation to cryo-storage of the produced O\textsubscript{2} and H\textsubscript{2} (see Section~\ref{sec:isru_model} for a detailed explanation of the models)\revaa{, except for power and water purification subsystems}. 
\revaa{Power system is not considered in this paper for the sake of simplicity of analysis. Choosing an appropriate power generation system (e.g., photovoltaic, nuclear fission or radioisotope thermoelectric generator) as well as an energy storage system (e.g., battery or fuel cell) requires a large number of assumptions and can add significant complexity in the study. Furthermore, these systems can be shared with crewed a lunar base as well, which is out of the scope of this paper. The interested reader is referred to Refs.~\cite{Chen.2020, Kaczmarzyk.2021} for power and energy storage systems on the Moon.
It should be noted that power systems can play a critical role in the mass estimation of ISRU systems, and it should be assessed in the future.}
The infrastructure and operations to further utilize these products are not addressed in this work.

Margins are included in the mass budgets following the methodology of the American Institute of Aeronautics and Astronautics Mass Properties Standard \citep{AIAA.2015}. This standard recommends a minimum increase of 30\% of the estimated subsystem mass to include a growth margin as well as an additional margin that accounts for uncertain packaging and structural or neglected components. Similarly, a 30\% margin is considered for the power budget. 

Table \ref{tab:assumptions} collects the global assumptions considered in this study, including the radiative outer space and average surface temperatures for each site, which are used for the heat loss and radiative heat exchange calculations, and the dry and icy regolith density dependencies on depth and water ice content. Individual assumptions are described for each model in Section~\ref{sec:isru_model}. Moreover, Section~{\ref{sec:uncertainty_in_lunar_isru}} discusses the values of the following uncertain parameters considered in this study: feedstock particle size, silicate and water ice content in the regolith, \revaa{WE} efficiency, and ISRU plant operational availability.

\begin{table}[t]
    \centering
    \caption{Global study assumptions.}
    \label{tab:assumptions}
    \begin{tabularx}{\linewidth}{>{\raggedleft}X  >{\raggedright\arraybackslash}X} 
    \hline\hline
        Required production rate (baseline)  & 10 t/a of O\textsubscript{2} and 1.25 t/a of H\textsubscript{2}\\
        PEL/PSR traverse distance                & 5000 m \\
        Mass margin                              & 30\% \\
        Power margin                             & 30\%	\\
        PEL avg. temperature                     & 200 K\\
        PSR avg. temperature                     & 80 K\\
        Outer space temperature                  & 4 K\\
        \revaa{Avg. silica content}              & \revaa{45 wt\%}\\
        \revaa{Avg. water ice content}           & \revaa{4.2 wt\%}\\
        Dry regolith density\textsuperscript{1} & \revaasecond{$\rho_\mathrm{dry}(z~ [\mathrm{m}]) = 1800-700\cdot\exp(-z/0.06) ~[\mathrm{kg}/ \mathrm{m}^3]$}\\
        Icy regolith density\textsuperscript{2} &\revaasecond{$\rho_\mathrm{icy}(w_\mathrm{ice}~[\mathrm{wt\%}]) = \rho_\mathrm{dry}\cdot(1 + w_\mathrm{ice}) ~[\mathrm{kg}/ \mathrm{m}^3]$}\\\hline\hline
    \end{tabularx}
    \raggedright
    \footnotesize
    \textsuperscript{1} \revaasecond{The regolith depth in meters is denoted by $z$}, with $z = 0~ [m]$ at the lunar surface. Based on Hayne et al.~\cite{Hayne.2017}. \\
    \textsuperscript{2} The icy regolith density is calculated assuming that water ice mass fraction (\revaasecond{$w_\mathrm{ice}$}) accumulates uniformly within the regolith pores.
\end{table}
\subsection{Technology selection} \label{sec:technology_selection} 
Over the last decades, numerous strategies for extracting oxygen from dry lunar regolith have been proposed in the literature \citep{Taylor.1992b, Schluter.2020, Schwandt.2012}. 
From these strategies, ilmenite reduction by hydrogen followed by the electrolysis of water and carbon-based reductants (e.g., CO, CH\textsubscript{4}, C) has been considered as feasible and viable \revb{options \cite{Schwandt.2012}}. \revaa{CR} of partially molten regolith, with water electrolysis and CH\textsubscript{4}-reforming, the direct electrochemical reduction of molten regolith, the electrolysis in a bath of molten salts, or vacuum thermal decomposition of lunar soils, have also been extensively researched \citep{Schluter.2020}.

CR of partially molten regolith is chosen as O\textsubscript{2} extraction technology in this study due to its high yield, as it can reduce most regolith minerals, including the most prevalent silicates.
Hydrogen reduction is excluded from further consideration since hydrogen is mostly limited to reducing iron-bearing minerals such as ilmenite, which are \rev{expected to} only present in about 0.5 wt.\% of the lunar \rev{Sothern polar highland regolith \cite{Taylor.2010}}. As demonstrated by Guerrero-Gonzalez and Zabel~\cite{GuerreroGonzalez.2023}, the low ilmenite content significantly increases the total hardware mass of the plant, highlighting its ineffectiveness as an oxygen extraction technology in these polar regions.

Moreover, other O\textsubscript{2} extraction technologies, such as electrochemical reduction or vacuum thermal decomposition, do not produce water but oxygen directly from the regolith. Therefore, \revb{selecting either of these technologies could lead to larger total system mass and power due to the lack of a shared H\textsubscript{2}O electrolysis infrastructure.}

The extraction of water (or hydroxyls) can be done by heating lunar icy regolith until volatiles vaporize or desorb, followed by capturing the released gases \citep{Schluter.2020}. Generally, heat can be applied to the icy regolith by \revb{electric heating} or concentrated sunlight. Note that an alternative approach of using microwave heating also has gathered attention for its potentially higher extraction efficiency\cite{Cole.2023}, which should be further discussed in the future. The process of thermal extraction relies on the low-pressure environment on the Moon for sublimation to occur. By avoiding the liquid phase of water, the produced water vapor can outgass and be captured.

The main two principles of operation include in-situ water extraction through direct surface heating \citep{Sowers.2019, Purrington.2022, Schieber.2022} or drill rods \citep{He.2021, Liu.2023}, and excavation and transportation of icy regolith into an enclosed system followed by subsequent heating. This heating process can be continuous \citep{Collins.2023} or in batches \citep{Kiewiet.2022}. Kiewiet et al.~\cite{Kiewiet.2022} traded off in-situ against excavated water extraction methods, with the excavated designs generally scoring significantly better than the in-situ ones. Therefore, in this study, an excavator delivers icy regolith to a water extraction plant, where water is thermally extracted in crucibles.

\subsection{Concept of operations}\label{sec:concept_operations}
\subsubsection{\rev{Parallel hybrid architecture}}\label{subsec:conops_ph}
Six systems are involved in the overall concept of operations of the \revaa{PH} ISRU production architecture: (1) a dry-regolith mining (excavation and disposal) system, (2) a dry-regolith processing plant, (3) an H\textsubscript{2}/O\textsubscript{2} production and storage plant, (4) an icy-regolith mining (excavation and disposal) system, (5) an icy-regolith processing plant, and (6) a water transportation system between both ISRU plants, as depicted in Fig.~\ref{fig:type_a}.

At \revaa{a} PEL, an excavator delivers dry regolith to the processing plant to extract oxygen. To avoid disturbances, the excavation site is located 100 m away from the ISRU facility~\cite{Linne.2021}. After delivering a fresh regolith batch, the same vehicle is loaded with ISRU-produced tailings and drives first to a disposal site, also located 100 m away, before going back to the excavation site. The dry-regolith excavator recharges at the dry-regolith processing plant.

At \revaa{a} PSR, a similar scenario is defined. Another excavator delivers icy regolith to the icy-regolith processing plant, collects dry processed regolith, and dumps it at a disposal site before going back to the icy-regolith excavation site. The three locations are also positioned 100 m away from each other. The number of excavators for the dry- and icy-regolith is calculated based on the processing times, travel times, and regolith throughputs needed. The icy-regolith excavator only recharges at the icy-regolith processing plant without leaving the PSR.

A similar mobile water transportation solution to the one discussed by Kleinhenz and Paz~\cite{Kleinhenz.2020} is chosen to allow for a fair comparison between different system architectures.
A tanker remains at the icy-regolith processing plant to directly capture water into its tank, while others deliver H\textsubscript{2}O to the H\textsubscript{2}/O\textsubscript{2} production plant and come back to the PSR. 
This approach saves hardware mass since water vapor is allowed to freeze directly inside the tank, avoiding the need for additional cold traps. Moreover, due to the low PSR temperatures, no power is required to maintain the state of water ice. Once the tanker leaves the PSR, water starts to thaw due to solar radiation. 
The water tankers recharge at the H\textsubscript{2}/O\textsubscript{2} production and icy-regolith processing plants. \revb{The total number of the water tankers is selected from a range of two to 10 to minimize the total landed mass of the entire system. This number affects travel times, tanker capacity, water throughputs, and processing times needed.}

Liquid water delivered to the H\textsubscript{2}/O\textsubscript{2} production plant is subsequently electrolyzed into hydrogen and oxygen. Together with the oxygen extracted from dry regolith, they are liquefied and cryogenically stored.

\subsubsection{\rev{Series hybrid architecture}}
The \revaa{SH} architecture (Fig.~\ref{fig:type_b}) involves placing both the dry- and icy-regolith processing plants within the PSR to further process dry regolith after water extraction to obtain oxygen. Water extracted from icy regolith and generated through carbothermal reduction is transported to a PEL for electrolysis, liquefaction and storage, requiring the same water tanker system as the \revaa{PH} architecture.

\rev{By processing excavated icy regolith utilizing multiple technologies, this architecture can potentially achieve the highest oxygen yield per unit mass of excavated regolith, leading to slower excavation rate.}
However, significant operational \rev{and technological} challenges \rev{also} arise in this \rev{architecture}. Colder surface temperatures would negatively affect high-temperature ISRU processes, increasing energy consumption where no direct sunlight as a power source is available.

\section{Lunar ISRU hybrid plant modeling}\label{sec:isru_model}
Fig. \ref{fig:architecture} shows the \revaa{flow diagrams of the }system architectures \revaa{studied in this paper}. For \revaa{several} subsystems,  \revaa{technologies are} modeled using ISRULib. ISRULib is an open-source component- and system-level library of ISRU models developed by the Technical University of Munich to carry out high-level technological trade-offs and preliminary architectural definitions \citep{GuerreroGonzalez.2023b}.
To estimate the mass and power of the hybrid architectures proposed in this paper, parametric sizing models for carbothermal reduction and water extraction subsystems are newly developed.
The models estimate mass and power budgets, as well as the performance of different technologies, based on analytical parametric calculations, \revaa{surrogate models from numerical simulations, and already existing ISRU hardware. Table \ref{tab:subsystem_assumptions} lists subsystem models employed in this paper. Each of them is explained below in detail.}

\begin{figure*}[bth!]
\centering
     \begin{subfigure}[b]{\textwidth}
         \centering
         \includegraphics[width=0.75\linewidth]{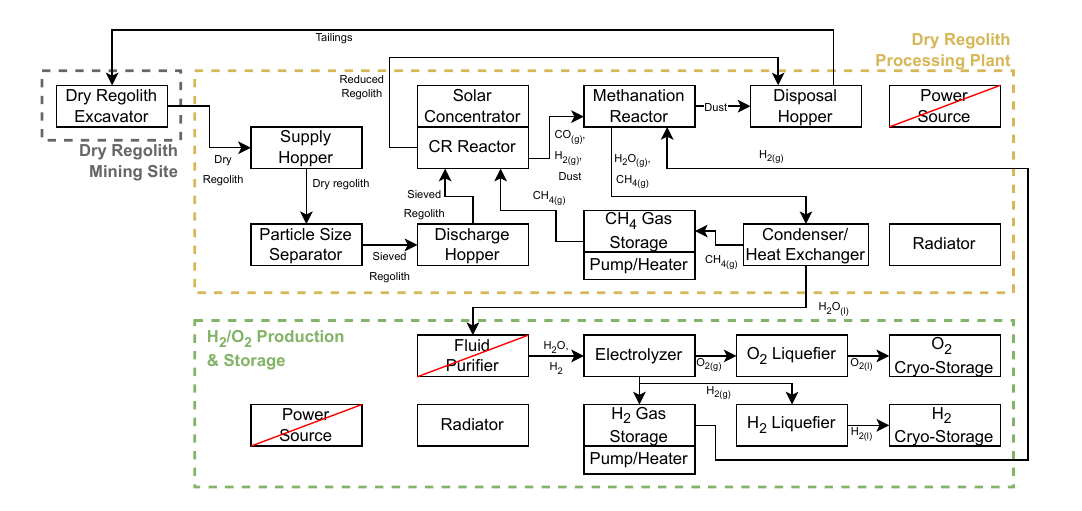}
         \caption{\revaa{Carbothermal reduction architecture.}}
         \label{fig:cr_system}
     \end{subfigure}\\
     \vspace{0.25 cm}
     \begin{subfigure}[b]{\textwidth}
         \centering
         \includegraphics[width=0.75\linewidth]{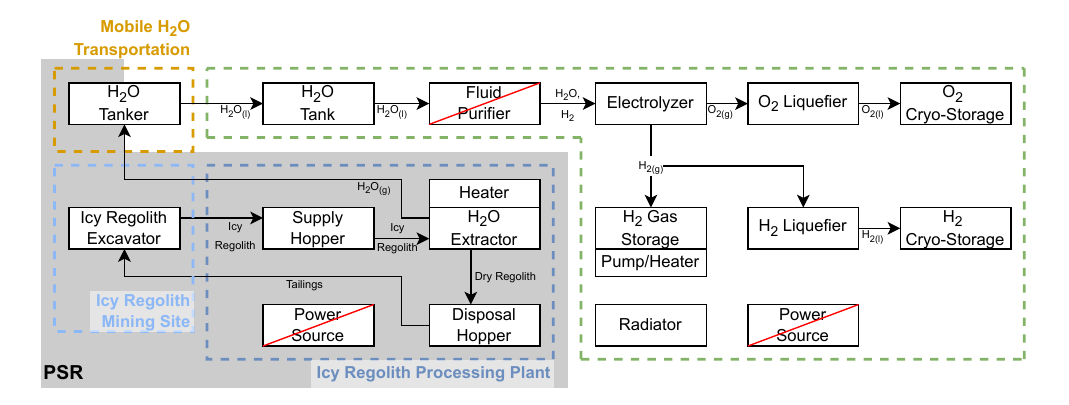}
         \caption{\revaa{Direct water extraction architecture.}}
         \label{fig:we_system}
     \end{subfigure}
     
\caption{\revaa{Flow diagrams of ISRU architectures considered in this paper. Mass flows across subsystems are shown. The diagram blocks crossed out in red are not modeled.}}
\end{figure*}

\begin{figure*}[bth!]\ContinuedFloat
\centering
     \begin{subfigure}[b]{\textwidth}
         \centering
         \includegraphics[width=0.75\linewidth]{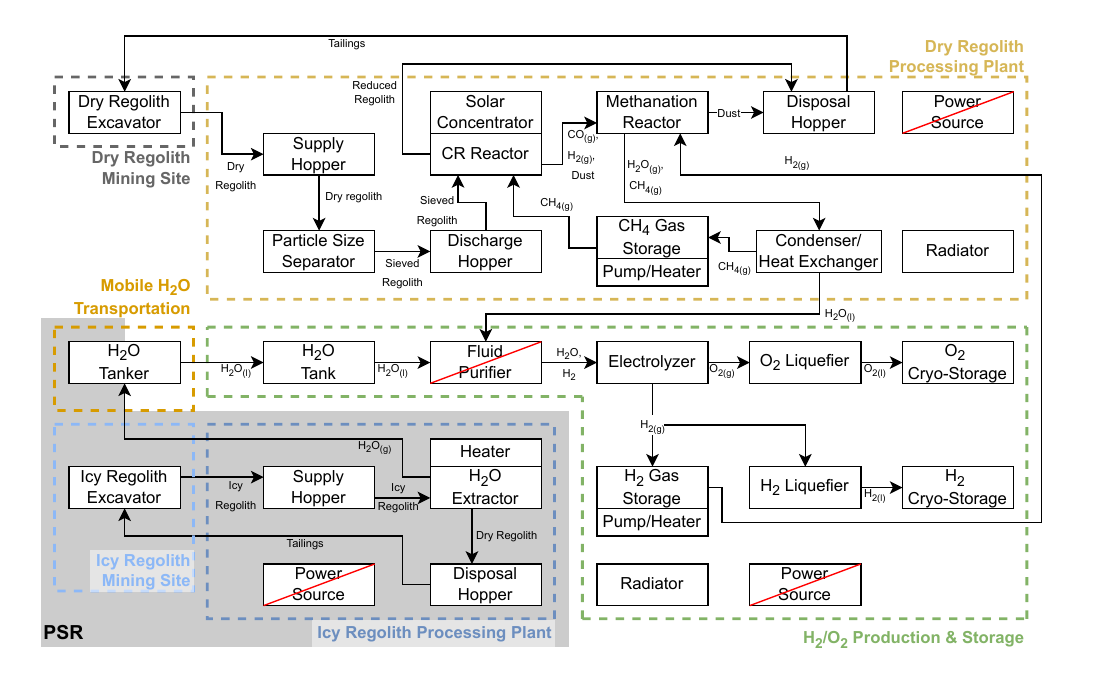}
         \caption{\rev{Parallel hybrid architecture.}}
         \label{fig:type_a_system}
     \end{subfigure}\\
     \vspace{0.25 cm}
     \begin{subfigure}[b]{\textwidth}
         \centering
         \includegraphics[width=0.75\linewidth]{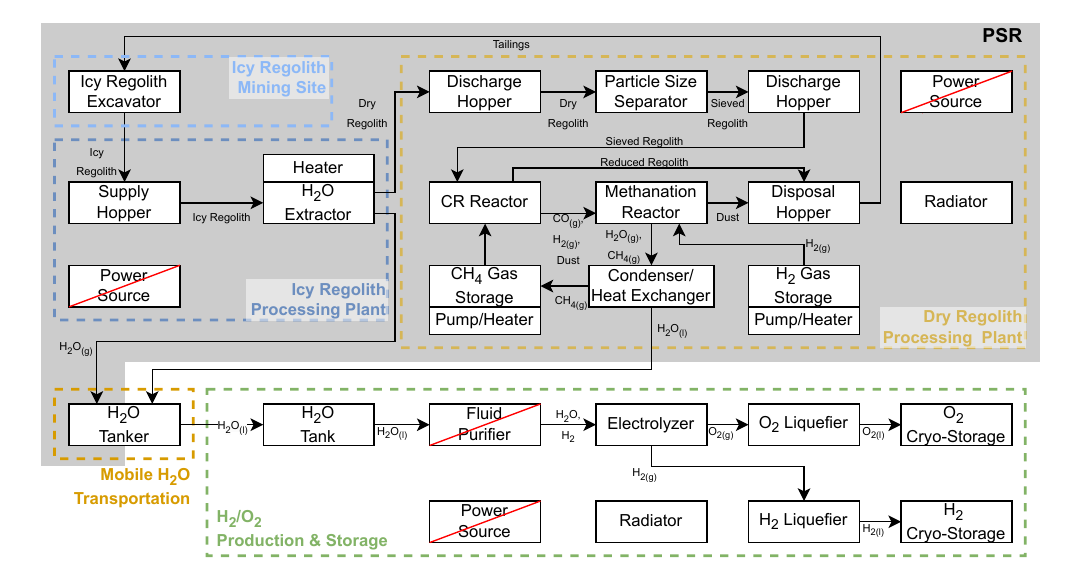}
         \caption{\rev{Series hybrid architecture.}}
         \label{fig:type_b_system}
     \end{subfigure}
\caption{\revaa{(Continued.)}}
\label{fig:architecture}
\end{figure*}

\begin{table*}[]
\centering
\caption{\revaa{Summary of subsystem model types and assumptions}}
\label{tab:subsystem_assumptions}
\resizebox{\textwidth}{!}{%
\revaa{
\begin{tabular}{llll}
\hline  \hline  
Element              & Model type & Notes and assumptions & \revaa{Refs.}\\ \hline
\textbf{Regolith mining} & &&\\
\quad Excavator & Existing hardware& RASSOR 2.0&\citep{Mueller.2016}\\
\textbf{Dry regolith processing} &&&\\
\quad CR reactor  & Param. size modeling& Material: Inconel& \cite{Balasubramaniam.2010, WoodsRobinson.2019, Keihm.1984}\\
\quad Methanation reactor& Specific mass& Reference: produced H\textsubscript{2}O, kg/h  &\cite{Schrenk.2015}\\
\quad Particle size separator &Param. size modeling&Material: aluminum& \cite{Benedi.2022}\\
\quad Condensor heat exchanger&Param. size modeling&Material: titanium& \cite{Schuster.2023}\\
\quad Hopper&Param. size modeling& Material: Inconel (discharge and disposal)&\\
\quad && Material: aluminum (supply)&\\
\textbf{Icy regolith processing}&&&\\
\quad H\textsubscript{2}O extractor&Surrogate
modelling from (Kiewiet et al.) &Material: Inconel&\cite{Kiewiet.2022}\\
&and param. size modeling&&\\
\quad H\textsubscript{2}O tank&Param. size modeling&Material: aluminum&\\
\quad Hopper &Param. size modeling&Material: aluminum&\\
\textbf{Mobile  H\textsubscript{2}O transportation} &&&\\
\quad H\textsubscript{2}O tanker&Param. size modeling& Based on payload ratio, aluminum tank,&\\
&&  and energy density&\\
\textbf{Electrolysis} &&       &\\
\quad PEM electrolysis stack&Param. size modeling&&\cite{Schrenk.2015}\\
\quad SOXE electrolysis stack&Param. size modeling&&\cite{Hinterman.2022}\\
\quad Dryers& Param. size modeling&Material: titanium&\cite{Schrenk.2015}\\
\quad Heat exchanger&Param. size modeling&Material: titanium&\cite{Schuster.2023}\\
\quad Pump& Specific mass & Reference: fluid flow rate, L/min&\cite{Micropump.2024}\\
\textbf{H\textsubscript{2}/ O\textsubscript{2} liquefaction} &&&\\
\quad H\textsubscript{2} Liquefaction subsystem  &Existing hardware, & Compressor (Deserranno, et al.)& \cite{Deserranno.2014}\\
& and param. size modeling& material: aluminum and neon& \cite{Schuster.2023}\\
\quad O\textsubscript{2} Liquefaction subsystem  &Existing hardware, & Compressor (CryoTel\circledR DS 30)& \cite{Cryotel.2024}\\
& and param. size modeling& material: aluminum and neon& \cite{Schuster.2023}\\
\textbf{Recycle and storage} &&&\\
\quad Gas storage tank&Param. size modeling&Material: carbon/epoxy&\\
\quad Cryogenic storage tank&Param. size modeling& Material: aluminum and MLIs&\cite{Johnson.2018}\\
\quad Pump&Specific mass&Same as above& -\\
\textbf{Thermal management} &&&\\
\quad Radiators & Specific mass&Reference: reject heat, kg/kW&\cite{GuerreroGonzalez.2023,Simonsen.1992}\\
\hline\hline
\multicolumn{4}{r}{param. = parametric}\\
\end{tabular}}}
\end{table*}

\subsection{Carbothermal reduction of dry regolith} \label{sec:carbothermal_reduction}
Carbothermal reduction (CR) is the process of reducing partially molten (often metal) oxides using carbon in a form, such as methane, as a reducing agent. This process generates carbon monoxide, which can be further processed to reform CH\textsubscript{4} and produce oxygen via water electrolysis. Historically, the CR process followed by methanation has been researched \citep{Linne.2021, Rice.1996, Troisi.2022} and demonstrated extensively to extract oxygen from lunar regolith simulants \citep{White.2023, Prinetto.2023}.

The chemical process can be generally expressed as:
\begin{align}
\mathrm{CH_4}&\rightarrow \mathrm{C} + \mathrm{2H_2} \label{eq:methane2carbon}\\
\mathrm{MeO}_x + x\mathrm{C}&\rightarrow \mathrm{Me} + x\mathrm{CO} \\
\mathrm{CO + 3 H_2} &\rightarrow \mathrm{CH_4 + H_2O} \label{eq:methanation}
\end{align}
where $\mathrm{MeO}_x$ and $\mathrm{Me}$ represent a general metal oxide and metal, respectively. CR can process the silicate minerals and iron-bearing minerals in lunar regolith \citep{Rice.1996}.
\revb{Gustafson et al.~\cite{Gustafson.2005} report that because of the widespread distribution of suitable metal oxides for CR on the lunar surface, mineral enrichment is not likely required for CR. However, it is still necessary to classify as-mined regolith by size to improve the reaction efficiency~\cite{Samouhos.2022}.
In this paper, it is assumed that the size-sorting beneficiation (see Section \ref{sec:beneficiation}) is applied before the CR process.}

\revb{Silica ({SiO\textsubscript{2}}) is the most abundant metal oxide in highland regolith, and is the focus of this work. While CR is capable of reducing various metal oxides at different temperatures, capturing this capability in the model adds considerable complexity with diminishing returns on fidelity.}

\revb{The reaction of silica with methane has been} extensively modeled by Balasubramaniam et al.~\citep{Balasubramaniam.2010} The reaction
\begin{align}
\label{eq:sio2_reaction}
\mathrm{SiO_2} + \mathrm{3C}&\rightarrow \mathrm{SiC} + \mathrm{2CO}
\end{align}
occurs at 1250-2000 $^{\circ}$C \citep{Lee.2010}. 
To achieve such high temperatures, \revaa{the use of solar thermal energy to melt a concentrated area of regolith has been proposed and demonstrated~\cite{Rice.2001, Gustafson.2010}.}
These studies also suggested to use regolith as a thermal insulator to protect reactor walls from molten regolith. The use of concentrated solar thermal energy has been experimentally simulated using lasers to demonstrate the CR process of lunar regolith simulants in vacuum conditions \revaa{as well} \citep{White.2023}.

\begin{figure*}[bth!]
\centering
\includegraphics[width=0.68\linewidth]{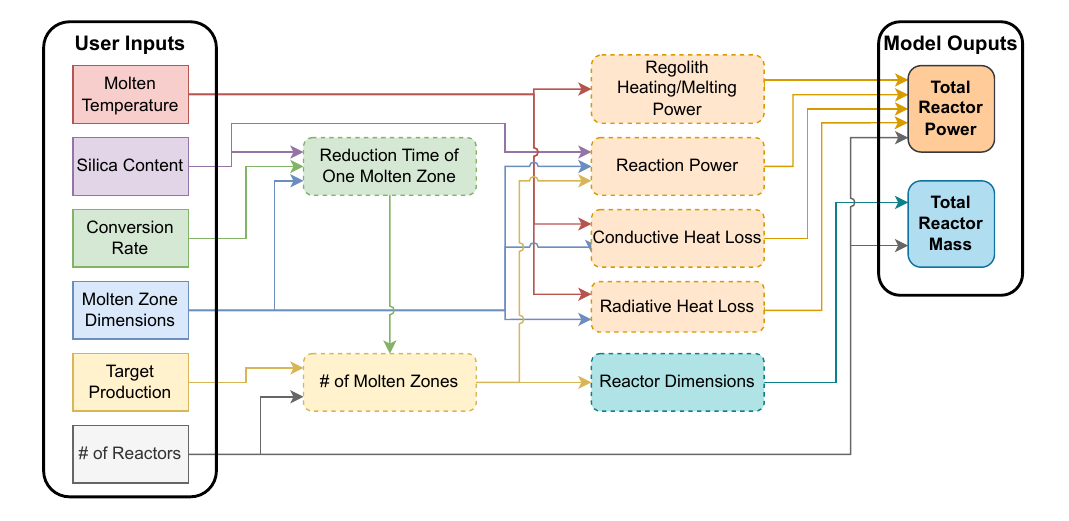}
\caption{High-level schematic of a parametric sizing CR reactor model.}
\label{fig:cr_model}
\end{figure*}

\begin{figure}[bth!]
\centering
\includegraphics[width=0.85\linewidth]{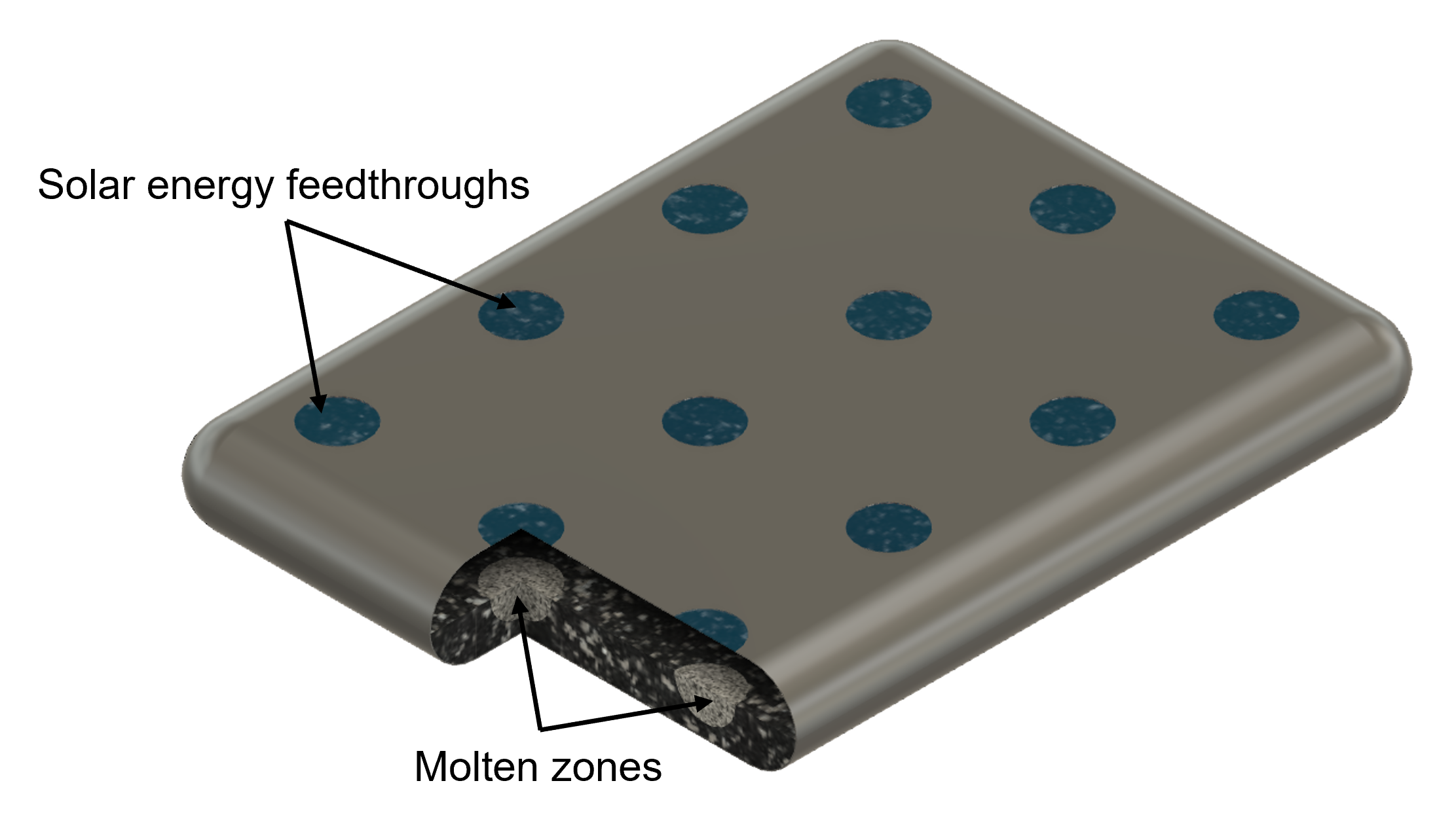}
\caption{CR reactor design. The number of molten areas is the same as the number of solar energy feedthroughs, which affects the dimensions of the reactor.}
\label{fig:cr_reactor}
\end{figure}

Figure \ref{fig:cr_model} depicts a high-level schematic of the parametric sizing model used for \rev{CR reactor} mass and power estimations. From silica content in the regolith, conversion rate, and dimensions of a molten regolith zone, the reaction time of one molten zone can be calculated using the model developed by Balasubramaniam et al.~\citep{Balasubramaniam.2010}
The estimated reaction time is used to calculate the minimum number of molten zones required to meet the target production rate. The number of molten zones can affect the dimensions of the reactor design, which affects the total mass of the reactor (see Fig.~\ref{fig:cr_reactor} for the design of the reactor).\rev{The main material of the reactor and its minimum allowable wall thickness is set to Inconel and 3.0 mm, respectively.}

The total power required from the solar concentrator can be divided into four: power for heating and melting the regolith, for carrying out the reaction, and the conductive and radiative heat losses to the environment. 
The heating power can be derived from the heat capacity of the regolith, which is modeled in the literature ~\cite{WoodsRobinson.2019, Keihm.1984}. The latent heat of lunar highland regolith is calculated as 478.6 kJ/kg by Schreiner et al.~\cite{Schreiner.2016}.
The enthalpy of \revb{the reactions described by Equations} \ref{eq:methane2carbon} and \ref{eq:sio2_reaction} are reportedly 75.6 kJ/mol and 689.8 kJ/mol, respectively \cite{Muscatello.2011}.
The conductive heat loss can be calculated from the geometry of the molten zone and the reactor, and the thermal conductivity of lunar regolith.
Cremers and Hsia \cite{Cremers.1974} reported the thermal conductivity of Apollo 16 highland regolith as $1.0-1.5 \times 10^{-3} \mathrm{W/ (m\cdot K)}$ at around 400 K. Due to this extremely low thermal conductivity of lunar regolith, the conductive heat loss can be practically ignored.
The radiative heat loss can be calculated from the molten zone geometry and its temperature and the environment temperature. Following the assumptions made in \cite{Muscatello.2011}, we use 0.7 as the emissivity of molten regolith, and we assumed that there is an insulation cover on each molten zone with a small hole on it. Due to these insulation covers, the radiation loss can be limited to 6.25\% of the case without the covers. With these assumptions, the heat loss from one molten zone can be calculated as about 26.6 W. \revaasecond{From these parameters and batch processing time, the average total power required for CR reactors is calculated.}

The mass of the methanation chamber is modeled based on the work of Schrenk~\cite{Schrenk.2015}. In this model, the mass of \rev{a Sabatier chamber is scaled based on a mass of those demonstrated in literature such as \cite{Junaedi.2014}. Following \cite{Schrenk.2015}, the specific mass used for the methanation subsystem is 17.9 kg per 1.0 kg/h of water produced.} \revaasecond{As discussed later in Section \ref{sec:result_deterministic_baseline}, the mass of the methanation chambers is not a main mass driver, and therefore, this simple specific mass model is enough considering the scope of this paper. Parametric sizing models for different chamber designs would be desirable for more detailed CR architecture design decision-making.}



\subsection{Water extraction from icy regolith} \label{sec:water_extraction}
\rev{In the proposed architecture,} water from icy regolith is thermally extracted in crucible water extractors \cite{Kiewiet.2022, Hegde.2012} in a PSR.
The performance of these crucible water extractors is modeled following the work of Kiewiet et al.~\cite{Kiewiet.2022}. 
\rev{Each extractor has} resistive heaters providing the required heating power. 
It is assumed that the entire extraction process, from icy-regolith excavation to water storage, has a 75\% mass recovery efficiency to account for line losses and losses occurring during excavation or vapor capturing. This value is in agreement with the assumptions made by Kleinhenz and Paz~\cite{Kleinhenz.2020}.

\rev{Kiewiet et al.~\cite{Kiewiet.2022} have optimized the heating time of a reactor with different input power and different water ice content to maximize the water yield using the COMSOL Multiphysics\textregistered ~software. In their study, the water content of 1, 5, and 15 weight \% (wt\%), and the input power of 500, 1500, and 2500 W were selected to see their effect on water extraction rate and energy efficiency.}

\begin{figure*}[bth!]
\centering
\includegraphics[width=0.68\linewidth]{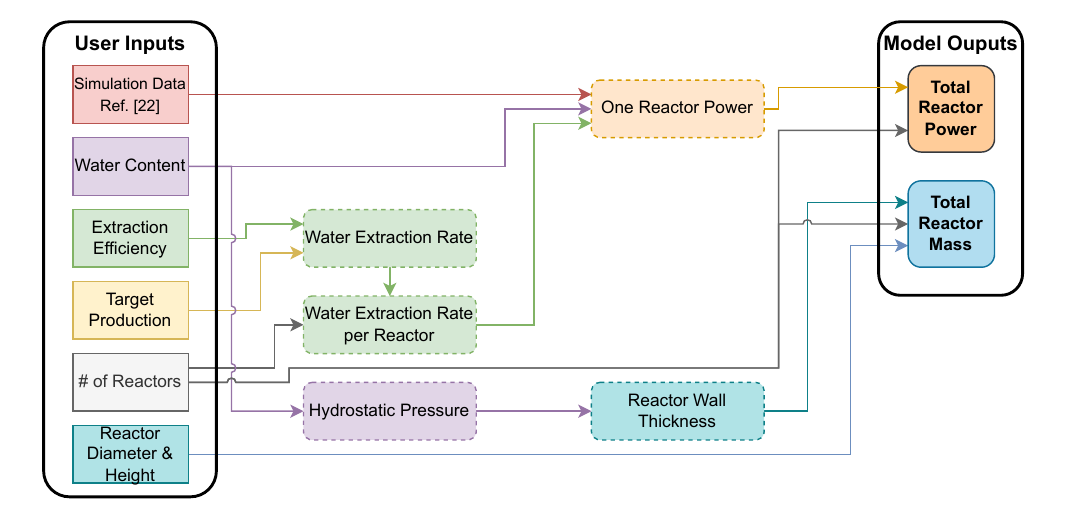}
\caption{\rev{High-level schematic of a parametric sizing WE reactor model.}}
\label{fig:we_model}
\end{figure*}

\rev{Figure \ref{fig:we_model} depicts a schematic of water extractors' mass and power estimation developed for this paper. From the extraction efficiency and the target oxygen and hydrogen production rate, the target water extraction rate can be calculated. From the relationship between the input power to an extractor, the average water ice content in excavated regolith, and the water extraction rate, mentioned above, the model can estimate the required power. Note that the relationship is fit to a quadratic equation from the data points from \revaa{Kiewiet et al.}~\cite{Kiewiet.2022}, and the maximum allowable power input per reactor is set to 2500 W to avoid extrapolation on the higher power side. When the input power requirement exceeds 2500 W, the number of reactors is modified iteratively.}

The mass of the reactor is estimated from the crucible reactor design with a diameter of 0.46 m and a height of 1.0 m~\cite{Kiewiet.2022}. \revb{Ignoring other volatile gases, }the required wall thickness is calculated from hydrostatic pressure based on the water content with a minimum thickness of 3.0 mm and Inconel is selected as reactor material.
\subsection{Supporting subsystems} \label{sec:supporting_subsystems}
Besides the principal oxygen and water extraction processes, the required supporting subsystems, including shared infrastructure, are also modeled to provide a holistic end-to-end representation of the entire ISRU production plants.
This subsection briefly summarizes the excavation, beneficiation and water transportation subsystems. For the shared subsystems, see Section.~\ref{sec:shared_infrastructure}.

\subsubsection{Excavation and handling subsystem} 
The excavator models are based on NASA’s Regolith Advanced Surface Systems Operations Robot (RASSOR) 2.0~\citep{Mueller.2016}. RASSOR 2.0 includes two sets of two bucket drums that produce opposing digging forces, causing a net-zero horizontal reaction force and enabling its operation under reduced lunar gravity. 
\revaa{It should be noted that excavation of icy regolith may be more challenging than that of dry regolith depending on how ice is formed in a PSR~\cite{Ricardo.2023}. Without having enough information about the water ice structure on the Moon, this paper assumes RASSOR 2.0 is used for both dry and icy regolith excavation. The required power for the excavation is also assumed to be the same between dry and icy regolith. While the first generation of RASSOR demonstrated the excavation of lunar regolith simulant with water ice~\cite{Mueller.2013}, further demonstration for various water ice content needs to be conducted.}

Both excavators deliver regolith to hoppers sized to accommodate three batches of the H\textsubscript{2}O extractor and CR reactor, respectively, to account for buffer storage due to operational mismatches. The discharge and disposal hoppers of the dry-regolith processing plant are made of Inconel to withstand the carbothermal reduction temperature. Since both supply hoppers and the disposal hopper of the icy-regolith processing plant have to withstand lower operating temperatures, aluminum is selected as a structural material in this case.

\subsubsection{Beneficiation subsystem} \label{sec:beneficiation}
A particle size separator is included to remove the coarse regolith fraction. Although the particle size effect on the carbothermal reduction of lunar regolith is not well-researched, Samouhos et al.~\cite{Samouhos.2022} claim that the smaller particle size leads to higher efficiency in the reduction by increasing the relative surface area between regolith and carbon. Haas and Khalafalla~\cite{Haas.1968} also reported a decreasing efficiency of carbothermal reduction around the silica particle size of 0.13 mm. Without further information on particle size effects on carbothermal reduction of lunar regolith, this paper assumes that the particle size separator removes particles with a diameter larger than 0.15 mm.
The particle size separator is based on the model by Linne et al.~\cite{Linne.2021}, where a trough covered by a size-sorting grate dumps regolith from a hopper to an auger conveyor that transports the granular material to the next subsystem. \revaa{The main mass and power driver of this beneficiation subsystem is the auger, which is modeled based on Benedi et al.~\cite{Benedi.2022}.} \revaasecond{Hence, while the size sorter in the study by Linne et al. is designed for removing particles larger than 25 mm, the differences in the mass and power consumption between their system and the proposed one are assumed trivial. Note that there may be more appropriate techniques for the size sorting of small highland regolith particles. For instance, Al Moinee et al.~\cite{AlMoinee.2024} used an electrostatic sieve to separate particles smaller than 0.5 mm into three groups. For more detailed ISRU design consideration, a tradeoff study of available beneficiation technologies and a careful consideration of feasible operations would be necessary.}

\subsubsection{Water transportation} 
The model developed by Kleinhenz and Paz~\cite{Kleinhenz.2020} is considered for the water tankers discussed in Section~\ref{sec:concept_operations}. The tankers are composed of a mobility platform and a payload, which includes a water tank, a battery, and a communication and navigation unit. The mobility platform mass is based on a payload ratio (payload mass/mobility platform mass) of 1.5, similar to RASSOR 2.0. The aluminum tank is modeled based on its water capacity, including a 50\% ullage. The battery energy density is 140 Wh/kg, allowing for a maximum 80\% discharge. These same battery properties are used for the RASSOR 2.0 excavators. The charging stations at the propellant production and H\textsubscript{2}O extraction plants can provide enough power to allow for a 5-hour excavator and a 10-hour tanker recharge time.
\revaa{As mentioned in Sec.~\ref{subsec:conops_ph}, water starts to thaw after a tanker leaves the PSR. The required time for this process is calculated from the thermal environment of the tanker. The developed model compares this time with the charge time and adds extra time for waiting for this process to be done when necessary.}

The tankers deliver water to a fixed aluminum tank at the H\textsubscript{2}/O\textsubscript{2} production plant, which is 3 times larger than the total tankers' water tank volume to account for operational mismatches. The water tank mass also includes a Commercial-Off-The-Shelf (COTS) pump \revb{, which is modeled based on those for space applications by Micropump, Inc.~\cite{Kleinhenz.2017, Schuster.2023, Micropump.2024}}

\subsection{Shared infrastructure}\label{sec:shared_infrastructure}
Most of the shared infrastructure between the oxygen and hydrogen coproduction from lunar icy and dry regolith is shown in green \revaa{dashed lines} in Fig.~\ref{fig:architecture}. This shared infrastructure includes the water tank and electrolyzer, the H\textsubscript{2} and O\textsubscript{2} liquefaction and cryostorage subsystems, and high-pressure tanks to supply gas for the carbothermal reduction, accounting for buffer storage due to operational mismatches. A COTS pump is included before the pressurized vessels to store H\textsubscript{2} and CH\textsubscript{4} at \rev{300 bar, and 200 bar, respectively \cite{Dottori.2023},} and average PEL temperature. 
It must be noted that the energy required to cool down the exhaust gas stream before the electrolysis is partially recuperated and used to preheat the intake methane stream. Therefore, an additional Condensing Heat Exchanger (CHX) is included between the pressurized vessel and the reactor to further increase the gas temperature to the operating temperature.
\revaa{Optimal storage conditions of reactants should be discussed in the future.}

It must be noted that no fluid purifier is considered in this study. The criteria for designing such a subsystem \revb{are largely undefined} since \revb{lunar} water contaminants are yet to be fully characterized \citep{Schluter.2021}.

\subsubsection{\revaa{Thermal management subsystem}}
\revaa{Some subsystems require thermal management to exhaust heat. In this paper, radiators are considered to accomplish this purpose. Subsystems that require radiators in this paper are a CHX for CR, an electrolysis subsystem, and a liquefaction subsystem.}

\revaa{Due to different thermal conditions between a PSR and a PEL, the required surface area to reject a unit amount of heat should be different. Guerrero-Gonzalez and Zabel~\cite{GuerreroGonzalez.2023} estimated 40 kg of radiator required to reject 1 kW heat based on Guerrero-Gonzalez~\cite{GuerreroGonzalez.2021} at a PSR. Simonsen et al.~\cite{Simonsen.1992} also estimated the required mass of a radiator system for different regions on the Moon. In their study, to reject 201 kW heat, the mass of the thermal management system was estimated as 10, 681 kg at the South Pole. Therefore, in this paper, the mass factor of 53 kg/kW is considered for radiators in a PEL.}

\revaa{Since CR is conducted in a PSR for the SH architecture, a radiator is assumed to be placed in a PSR. The rest of the radiators for the SH architecture as well as those of the PH architecture are placed in a PEL.}

\subsubsection{Electrolysis subsystem}\label{sec:electrolysis}

\begin{figure*}[pos=bth]
\centering
     \begin{subfigure}[t]{0.56\textwidth}
         \centering
         \includegraphics[width=\linewidth]{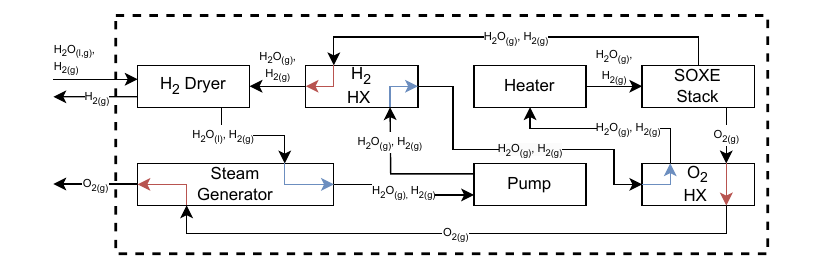}
         \caption{High-temperature Solid Oxide Electrolysis (SOXE). Inside the steam generator and Heat Exchangers (HXs), heat is transferred from the hot (red) to the cold (blue) fluid streams.}
         \label{fig:electrolysis_soxe}
     \end{subfigure}
    \begin{subfigure}[t]{0.43\textwidth}
         \centering
         \includegraphics[width=\linewidth]{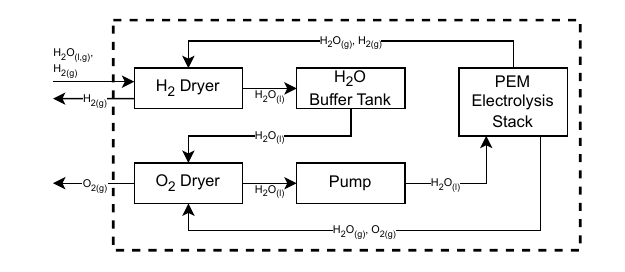}
         \caption{Proton Exchange Membrane (PEM) electrolysis.}
         \label{fig:electrolysis_pem}
     \end{subfigure}
\caption{Electrolysis subsystem (Electrolyzer) architectures. Mass flows across components are shown.}
\label{fig:electrolysis}
\end{figure*}

Different water electrolysis technologies, such as high-temperature Solid Oxide Electrolysis (SOXE)~\cite{Dickson.2021}, Proton Exchange Membrane (PEM) electrolysis~\cite{Chow.2018}, or alkaline electrolysis~\cite{Sakurai.2013}, have already been discussed in the context of space exploration \citep{Akay.2022}.
\revb{In this study, the two most commonly researched electrolysis architectures, SOXE and PEM (Fig.~\ref{fig:electrolysis}), are evaluated regarding their  subsystem mass and required power.}

\revb{Table~\ref{tab:pem_vs_soxe} lists the mass and power consumption estimated for both systems sized to generate 10 t of oxygen and 1.25 t of hydrogen per lunar year. The mass and power in this table are summations of all components shown in Fig.~\ref{fig:electrolysis} and radiators. As can be seen in Table \ref{tab:pem_vs_soxe}, the mass of the PEM electrolysis subsystem can be significantly lighter, \revaa{while it requires larger radiators.
Regarding required power, the SOXE electrolysis requires less power.} 
Therefore, this paper employed the SOXE subsystem for further analysis.}

\begin{table}[]
\centering
\caption{Mass and power comparison of water electrolysis technologies. Sized for 10 t of oxygen per lunar year.}
\label{tab:pem_vs_soxe}
\revaasecond{
\begin{tabular}{lrr}
\hline\hline
 & SOXE& PEM\\\hline
\textbf{Mass  {[}kg{]}} &\textbf{418}   &\textbf{567}\\
\quad Electrolysis &101  &12 \\
\quad Termal management &317  &555 \\\hline
\textbf{Power  {[}W{]}} &\textbf{7,756}   &\textbf{23,000}\\
\hline\hline
\end{tabular}}
\end{table}

\subsubsection{Gas liquefaction and storage subsystem}
Johnson et al.~\cite{Johnson.2018} compared several oxygen liquefaction methods for use on the Martian surface. From their analysis, the tube-on-tank cycle was adopted for both hydrogen and oxygen liquefaction in this study due to its performance and simple adaptation to the lunar environment. The system consists of a tank equipped with small tubes mounted around its external wall. The working fluid circulates through the tubes and is cooled by a series of cryocoolers. The cryogenic tanks are sized in aluminum to accommodate the entire yearly gas production. No additional power is included for storage maintenance \revb{because the system is a zero boil-off design.}

\section{Mass, power, and sizing results}\label{sec:result_deterministic}
In this section, the mass and power budgets of the hybrid ISRU plant are compared with more conventional plant architectures. The budgets were estimated using the model described in Section~\ref{sec:isru_model}.
\rev{The landed mass and required power of each architecture are compared following the baseline case (Section~\ref{sec:result_deterministic_baseline}) and sensitivity analyses are conducted in Section~\ref{sec:result_sensitivity}. The developed model is further compared with past studies in Appendix \ref{sec:mass_power_comparison}.}



\subsection{Comparison of resource extraction architecture}
\label{sec:result_deterministic_baseline}
To compare all architectures, each plant is designed to produce the same LOX amount. Since the hydrogen generated from the CR process is recycled for the methanation process (Equation \ref{eq:methanation}), additional LH\textsubscript{2} and its storage tank are assumed to be brought from Earth to compensate for the lack of produced hydrogen. Furthermore, even though both methane and hydrogen are recycled in the CR process, their recycling rate is unlikely to be 100\%. Lavagna et al.~\cite{Lavagna.2023} reported a recycling  rate of 91\% from their experiments. In this paper, we use this value of 91\% as a baseline recycle rate assuming additional \rev{gaseous} methane and hydrogen are brought to the Moon to compensate for this reactant loss.
Due to the additional methane and hydrogen from Earth, the comparable \rev{landed} masses on the Moon of the CR architecture, and all hybrid architectures depend on the operational period. Table \ref{tab:result_deterministic} lists the estimated mass and power of each subsystem of each plant architecture scaled to produce 10 t of LOX per lunar year with 1.25 t of LH\textsubscript{2} either produced or brought to the plant additionally. The \rev{Parallel Hybrid architecture} on this table is designed to produce 5.0 t of LOX from water generated from the methanation, and the rest from the water from the PSR.

\begin{table*}[pos=bt]
\centering
\caption{Mass and power estimation of each architecture. Sized for 10 t of oxygen and 1.25 t of hydrogen per lunar one year. The operational period is set to one year.}
\label{tab:result_deterministic}
\begin{tabular}{l|rrrr|rrrr}
\hline  \hline  
                     & \multicolumn{4}{c|}{Mass {[}kg{]}${}^{\ast 1}$} & \multicolumn{4}{c}{Power {[}W{]}${}^{\ast 1}$} \\
Element              & CR      & WE      &PH${}^{\ast 2}$ & SH  & CR       & WE      & PH${}^{\ast 2}$       & SH\\ \hline
\textbf{Dry regolith mining} & && &&& &&\\
\quad Excavator           &         111&      -&     111&     -& 633& -& 633&-\\
\textbf{Dry regolith processing} &         &         &       &       &          &         &       &\\
\quad CR reactor           &     \revaa{381}&     -&   \revaa{216}& \revaa{340}& \revaasecond{41,641}& -& \revaasecond{20,821}&\revaasecond{29,595}\\
\quad Methanation reactor  &      54&      -&     27&     42&   -& -& -&-\\
\quad \revaa{Particle size separator}  &\revaa{5}&      -& \revaa{2}&\revaa{2}&  \revaa{15}&  -&  \revaa{8}&\revaa{10}\\
\quad \revaa{Condensor heat exchanger}  &\revaa{4}&      -& \revaa{2}&\revaa{3}&  -&  -&  -&-\\
\quad Hoppers  &\revaa{31}&      -& \revaa{31}&\revaa{27}&  \revaa{15}&  -&  \revaa{8}&\revaa{10}\\
\textbf{Icy regolith mining} &         &         &       &       &          &         &       &\\
\quad Excavator           &     -&       111&     111&    111&-&  633& 633&633\\
\textbf{Icy regolith processing} &         &         &       &       &          &         &       &\\
\quad H\textsubscript{2}O extractor & -& 461& 230&115 &-&  25,918& 12,959&5,593\\
\quad \revaa{H\textsubscript{2}O tanks}  &-&\revaa{174}& \revaa{68}&\revaa{174}&  -& -&-&-\\
\quad \revaa{Hoppers}  &-&\revaa{47}& \revaa{31}&\revaa{20}&  -& -&-&-\\
\textbf{Mobile  H\textsubscript{2}O transportation} &         &         &       &       &          &         &       &\\
\quad H\textsubscript{2}O tanker & -& 2,557& 1,014&  2,557& -& 1,769&   673&1,769\\
\textbf{Electrolysis} &         &         &       &       &          &         &       &\\
\quad \revaa{SOXE electrolysis stacks}  &   \revaa{100}& \revaa{100}& \revaa{100}& \revaa{100}& \revaa{7,182}& \revaa{7,182}&  \revaa{7,182}  &\revaa{7,182}\\
\quad Dryers&   \revaa{1}&  \revaa{1}&  \revaa{1}&  \revaa{1}&  \revaa{-}& \revaa{-}&  \revaa{-}  &\revaa{-}\\
\quad Heat exchangers&   \revaa{0.2}& \revaa{0.2}& \revaa{0.2}& \revaa{0.2}&  \revaa{-}& \revaa{-}&  \revaa{-}  &\revaa{-}\\
\quad Pumps&   \revaa{0.4}& \revaa{0.4}& \revaa{0.4}& \revaa{0.4}& \revaa{574}& \revaa{574}&  \revaa{574}  &\revaa{574}\\
\textbf{H\textsubscript{2}/ O\textsubscript{2} liquefaction} &         &         &       &       &          &         &       &\\
\quad H\textsubscript{2} Liquefaction subsystem  &    \revaa{-}&  \revaa{1,665}&  \revaa{834}&   \revaa{475}& \revaa{-}& \revaa{22,113}&  \revaa{11,057}&\revaa{6318}\\
\quad O\textsubscript{2} Liquefaction subsystem  &    \revaa{179}&  \revaa{179}&  \revaa{179}&   \revaa{179}&  \revaa{5,616}& \revaa{5,616}&  \revaa{5,616}&\revaa{5,616}\\
\textbf{Recycle and storage} &         &         &       &       &          &         &       &\\
\quad CH\textsubscript{4} storage&   \revaa{27}&    \revaa{-}&   \revaa{27}&  \revaa{27}&  \revaa{-}&        \revaa{-}&         \revaa{-}&\revaa{-}\\
\quad CH\textsubscript{4} recycling pump&   \revaa{0.4}& \revaa{0}& \revaa{0.4}& \revaa{0.4}&  \revaa{67}& \revaa{-}& \revaa{67}& \revaa{67}\\
\quad H\textsubscript{2} storage&   \revaa{39}&    \revaa{757}&   \revaa{431}&  \revaa{231}& \revaa{-}& \revaa{-}&  \revaa{-}  &\revaa{-}\\
\quad H\textsubscript{2} recycling pump&   \revaa{0.4}& \revaa{0}& \revaa{0.4}& \revaa{0.4}&  \revaa{5}& \revaa{-}& \revaa{2}& \revaa{3}\\
\quad O\textsubscript{2} storage&   \revaa{804}&\revaa{804}&  \revaa{804}&  \revaa{804}& \revaa{-}& \revaa{-}&  \revaa{-}  &\revaa{-}\\
\revaa{\textbf{Thermal management}} &         &         &       &       &          &         &       &\\
\quad \revaa{Radiators at PEL} &   \revaa{513}&  \revaa{453}& \revaa{483}&  \revaa{439}&       -&      -&    -&-\\
\quad \revaa{Radiators at PSR} &   -&  -& -&  \revaa{45}&       -&      -&    -&-\\
\textbf{Additional materials from Earth} &         &         &       &       &          &         &       &\\
\quad Additional H\textsubscript{2}${}^{\ast 3}$ &   \revaa{1,591}&  -&  \revaa{795}&  \revaa{1,220}&       -&      -&    -&-\\
\quad Additional H\textsubscript{2} tanks  &  \revaa{571}&  -&  \revaa{306}&  \revaa{445}&       -&      -&    -&-\\
\quad Additional CH\textsubscript{4}${}^{\ast 4}$ &   \revaa{906}&  -&  \revaa{453}&  \revaa{696}&      -&     -&     -&-\\
\quad Additional CH\textsubscript{4} tanks &   \revaa{61}&  -&  \revaa{38}&  \revaa{51}&      -&     -&     -&-\\\hline\
\textbf{Total} &   \textbf{\revaa{5,378}} &	\textbf{\revaa{7,309}}& 	\textbf{\revaa{6,298}}&	\textbf{\revaa{8,106}}&       \textbf{\revaasecond{55,732}}  &  \textbf{\revaa{63,806}}      &  \textbf{\revaasecond{60,223}}     &\textbf{\revaasecond{57,361}} \\
\hline\hline
\multicolumn{9}{r}{$\ast 1 $: All values include 30\% growth~\cite{AIAA.2015} except for additional H\textsubscript{2} and CH\textsubscript{4} mass.}\\
\multicolumn{9}{r}{$\ast 2 $: 5 t of LOX is generated from CR and the rest 5 t from WE (1:1 ratio).}\\
\multicolumn{9}{r}{$\ast 3 $: Additional H\textsubscript{2} for reactant loss (91\% recycling efficiency) and product compensation.}\\
\multicolumn{9}{r}{$\ast 4 $: 91\% recycling efficiency.}
\end{tabular}
\end{table*}

\begin{figure*}[pos=bt]
\centering
     \begin{subfigure}[b]{\textwidth}
         \centering
         \includegraphics[width=0.7\linewidth]{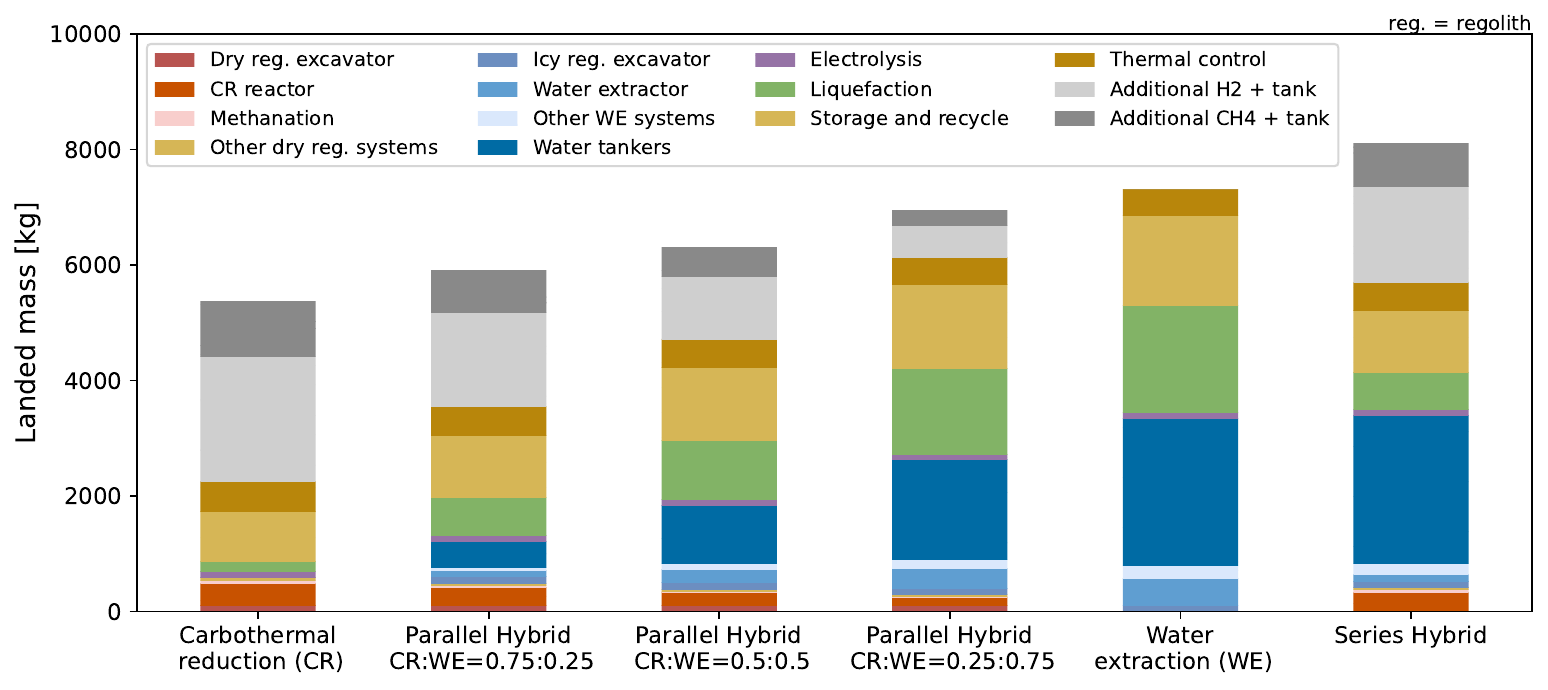}
         \caption{\revaasecond{One-year operation.}}
         \label{fig:sizing_mass_1}
     \end{subfigure}
     \vspace{0.25 cm}
     \begin{subfigure}[b]{\textwidth}
         \centering
         \includegraphics[width=0.7\linewidth]{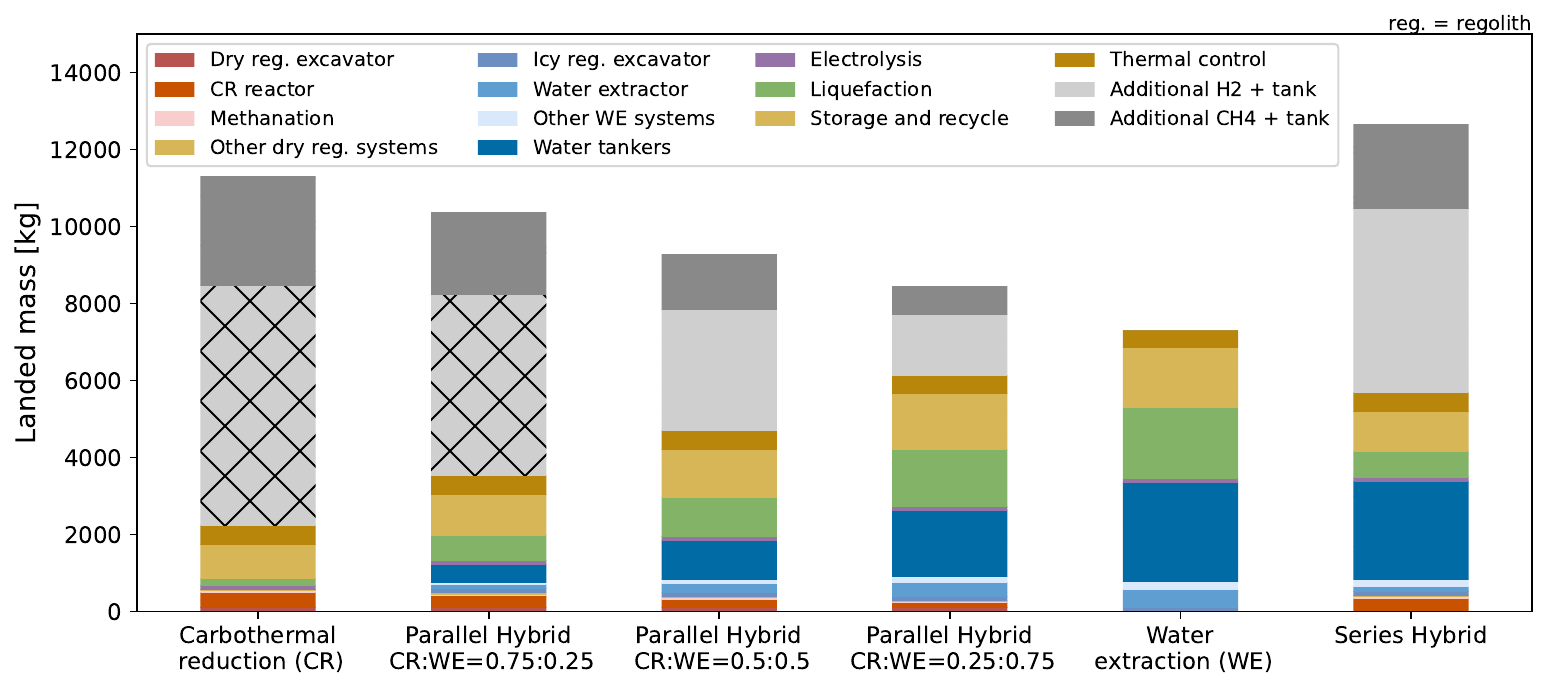}
         \caption{\revaasecond{Three-years operation.}}
         \label{fig:sizing_mass_3}
     \end{subfigure}
\caption{Mass comparison.}
\label{fig:sizing_mass}
\end{figure*}

\begin{figure*}[pos=bt]
\centering
\includegraphics[width=0.7\linewidth]{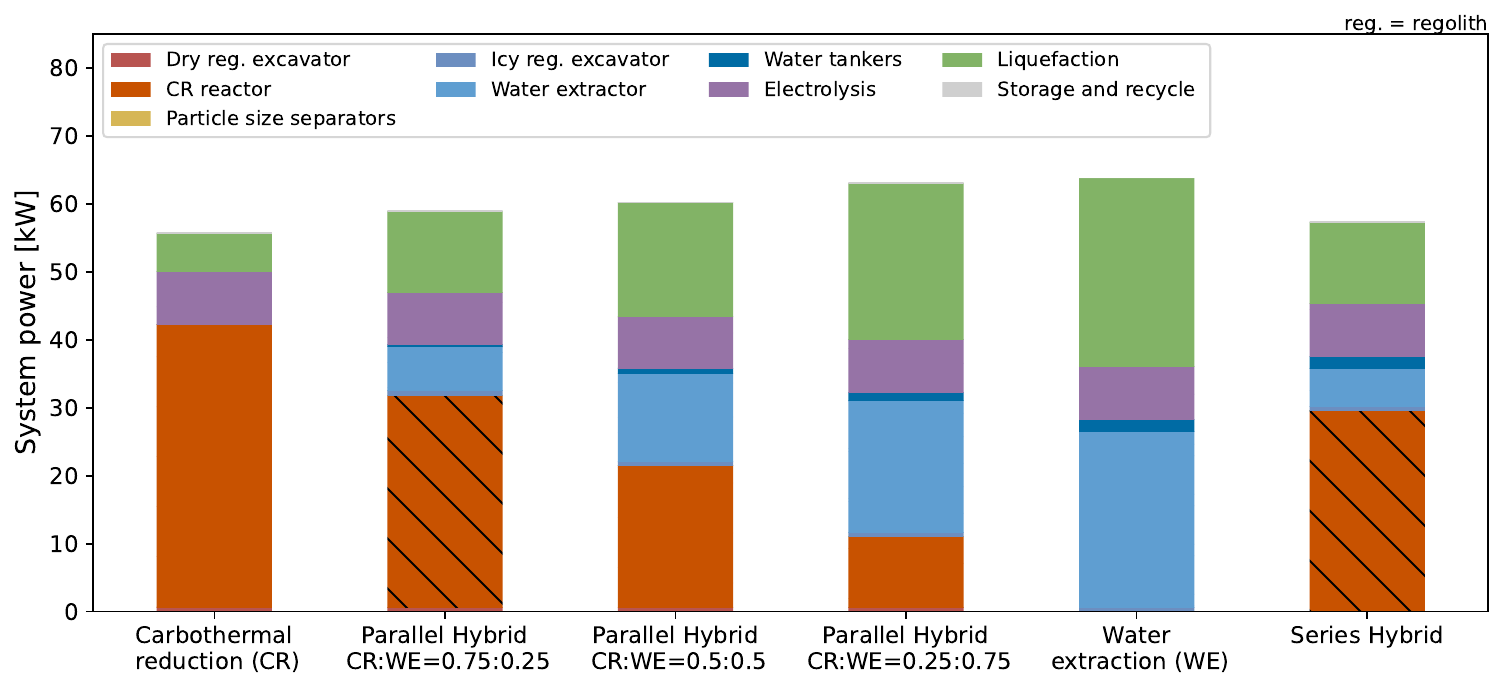}
\caption{\revaasecond{Power comparison.}}
\label{fig:sizing_power}
\end{figure*}

Figures \ref{fig:sizing_mass} and \ref{fig:sizing_power} are bar charts of mass and power breakdown. In addition to the four architectures in Table \ref{tab:result_deterministic}, the \rev{Parallel Hybrid architectures} with different hybrid ratios are added. The \rev{landed} mass of the CR architecture turned out to be the lightest compared to the others for one year of operation (Fig.~\ref{fig:sizing_mass_1}). As the operational period becomes longer, the additional hydrogen and methane from Earth become heavier, making the WE architecture the lightest (Fig.~\ref{fig:sizing_mass_3}). 
The main mass drivers other than additional methane and hydrogen are the water tankers and the liquefaction subsystem. Since liquefying hydrogen requires heavy cryocoolers (see, e.g., Deserranno et al.~\cite{Deserranno.2014}), the liquefaction subsystem of the CR architecture becomes the lightest.
The \revaa{SH} architecture turns out to be the heaviest in both analyzed operational periods. This is because the \revaa{SH} architecture largely depends on oxygen from CR rather than WE, requiring additional hydrogen and methane from Earth. Furthermore, since all water is produced in the PSR, this architecture requires the same amount of water tankers as the WE dedicated plant, which is another major mass driver.


\revaasecond{As can be seen in Table \ref{tab:result_deterministic} and Fig.~\ref{fig:sizing_power}, the main power drivers are the CR reactor, the water extractor, the electrolysis subsystem, and the liquefaction subsystem. As can be seen, the water extractor itself requires less power than the CR reactor. However, due to the liquefaction of hydrogen, the total difference in power consumption between the CR and WE architectures is about 8 kW. While the CR architecture requires the least power, the difference in power consumption between the CR and SH architectures is relatively small ($\sim 1.6$ kW). In the SH architecture, the low temperature at the PSR increases the energy required to heat regolith and leads to higher radiative heat loss. However, by partially relying on the less power-consuming WE process for oxygen production, the total power required for the CR reactor remains lower than that of the SH architecture.}

\revaasecond{It is important to note that the total power consumption is calculated under the assumption that water extraction and CR reactor operation occur simultaneously in hybrid architectures. If these processes are carried out at different times, power consumption at any given moment may be lower than the values presented here. Future studies should explore optimal operational strategies to minimize power consumption for each architecture.}

\subsection{Sensitivity Analyses}\label{sec:result_sensitivity}

\begin{figure}[pos=bth]
\centering
\includegraphics[width=0.95\columnwidth]{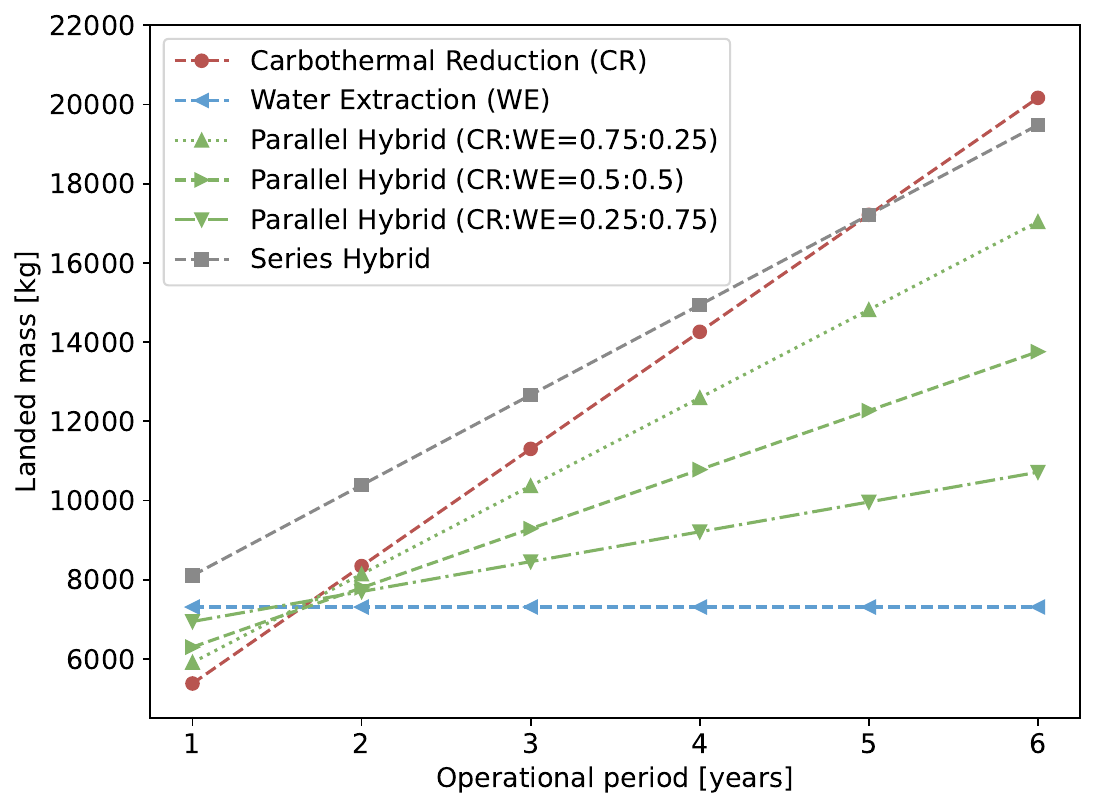}
\caption{\revaa{System mass as a function of operational period. The target production mass and the CR reactant recycle rate are set to 10.0 t of LOX and 1.25 t of LH\textsubscript{2}, and 91\%, respectively}}
\label{fig:sizing_mass_prod_period}
\end{figure}

\revaa{As mentioned, the CR process requires additional H\textsubscript{2} and CH\textsubscript{4}; therefore, the landed mass of any architecture with the CR process changes depending on the operational period. Figure \ref{fig:sizing_mass_prod_period} shows the change in the landed mass of each architecture in response to varying operational periods. Each architecture is designed to produce 10.0 t of LOX and 1.25 t of LH\textsubscript{2} annually. As can be seen in this figure, the order of architectures except for the \revaa{SH} is reversed between one year and two years' operation regarding their landed mass. For six years or longer operation, the CR architecture requires the most landed mass.}

\begin{figure}[pos=bth]
\centering
  \begin{subfigure}[b]{\columnwidth}{\includegraphics[width=0.95\linewidth,keepaspectratio]{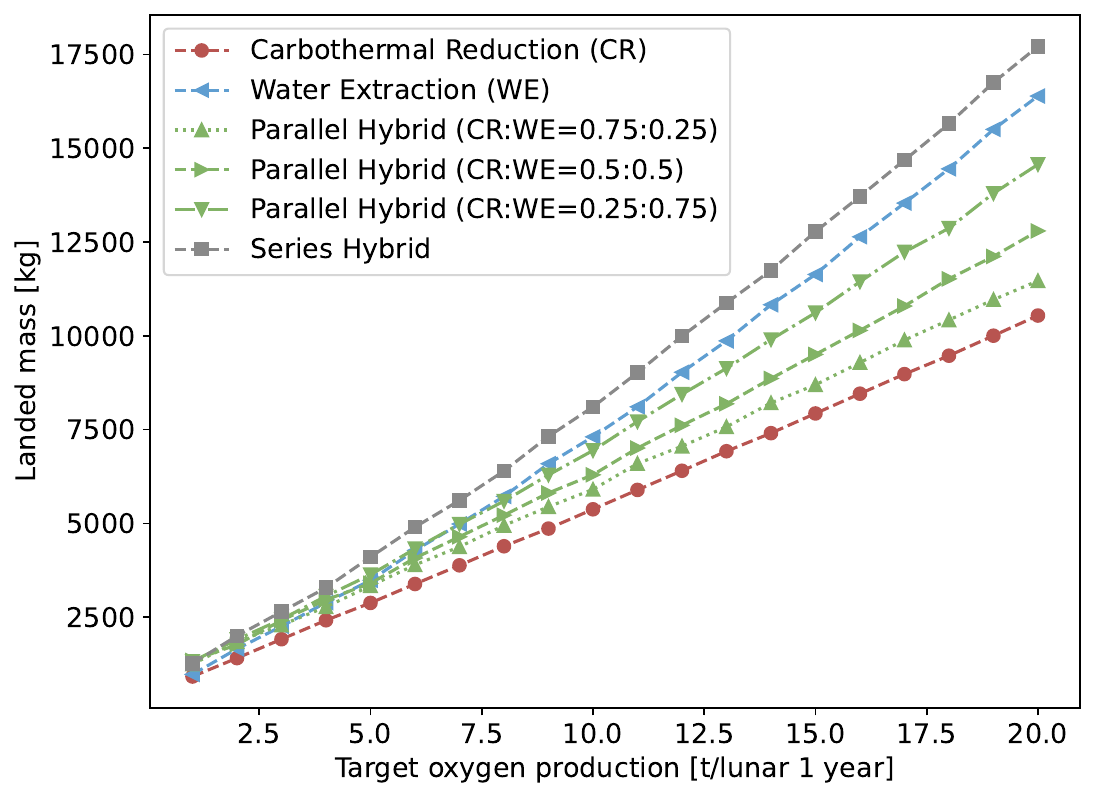}}
  \caption{\revaa{One-year operation.}}
  \label{fig:sizing_mass_target_ox_1} 
  \end{subfigure}\\
  \vspace{0.25 cm}
  \begin{subfigure}[b]{\columnwidth}{\includegraphics[width=0.95\linewidth,keepaspectratio]{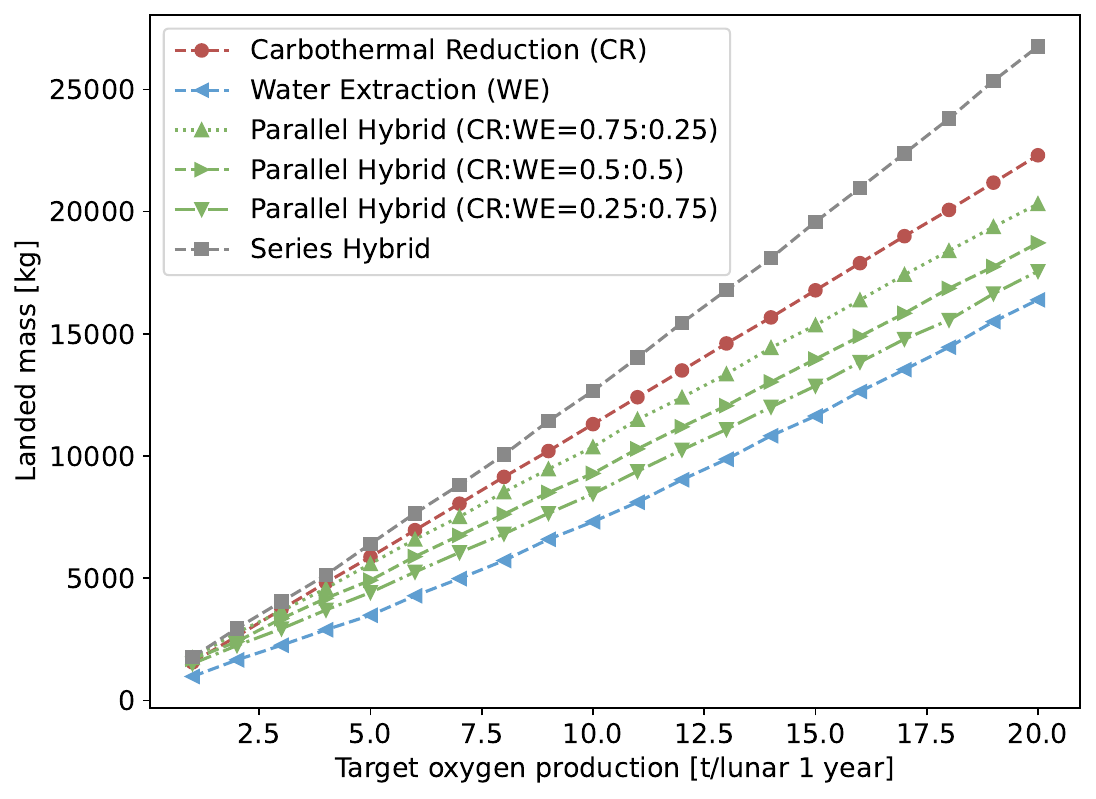}}
  \caption{\revaa{Three-years operation.}}
  \label{fig:sizing_mass_target_ox_3}
  \end{subfigure}
\caption{System mass as a function of target production rate. The CR Reactant recycle rate is set as 91\%.}
\label{fig:sizing_mass_target_ox}
\end{figure}

\begin{figure}[pos=bth]
\centering
\includegraphics[width=0.95\columnwidth]{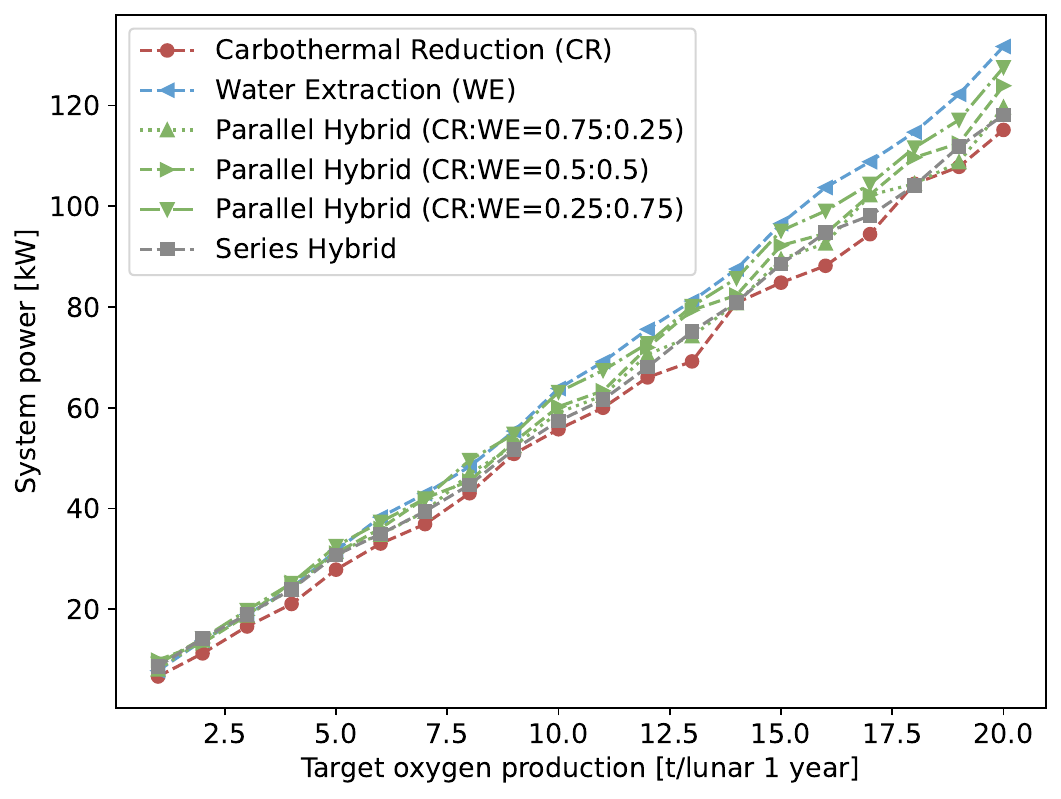}
\caption{\revaasecond{System power as a function of target production rate.}}
\label{fig:sizing_power_target_ox}
\end{figure}

The effect of the target production rate on the comparable mass and the power consumption are summarized in Figs.~\ref{fig:sizing_mass_target_ox}, and \ref{fig:sizing_power_target_ox}. As can be seen, there are near-linear trends for both mass and power against the production rate. 
Among the analyzed range, 1 – 20 \revaa{t} of oxygen per year (Fig.~\ref{fig:sizing_mass_target_ox_1}), the \revaa{SH} architecture is always the heaviest. While the \rev{CR architecture can be the lightest regardless of the target production for one year of operation (Fig.~\ref{fig:sizing_mass_target_ox_1})}, for three years of operation (Fig.~\ref{fig:sizing_mass_target_ox_3}), the advantage of the WE architecture becomes clearer due to the extra methane and hydrogen required for the other architectures.
When comparing the power consumption (Fig.~\ref{fig:sizing_power_target_ox}), the \revaasecond{CR} plant appeared to be the least power-consuming while the \revaasecond{WE} plant requires the most power \revaasecond{for the majority of the studied range. Compared to the landed mass, the variance in the power consumption among all architectures remains relatively small}.

\begin{figure}[pos=bth]
\centering
     \begin{subfigure}[]{0.99\columnwidth}
         \centering
         \includegraphics[width=0.95\linewidth]{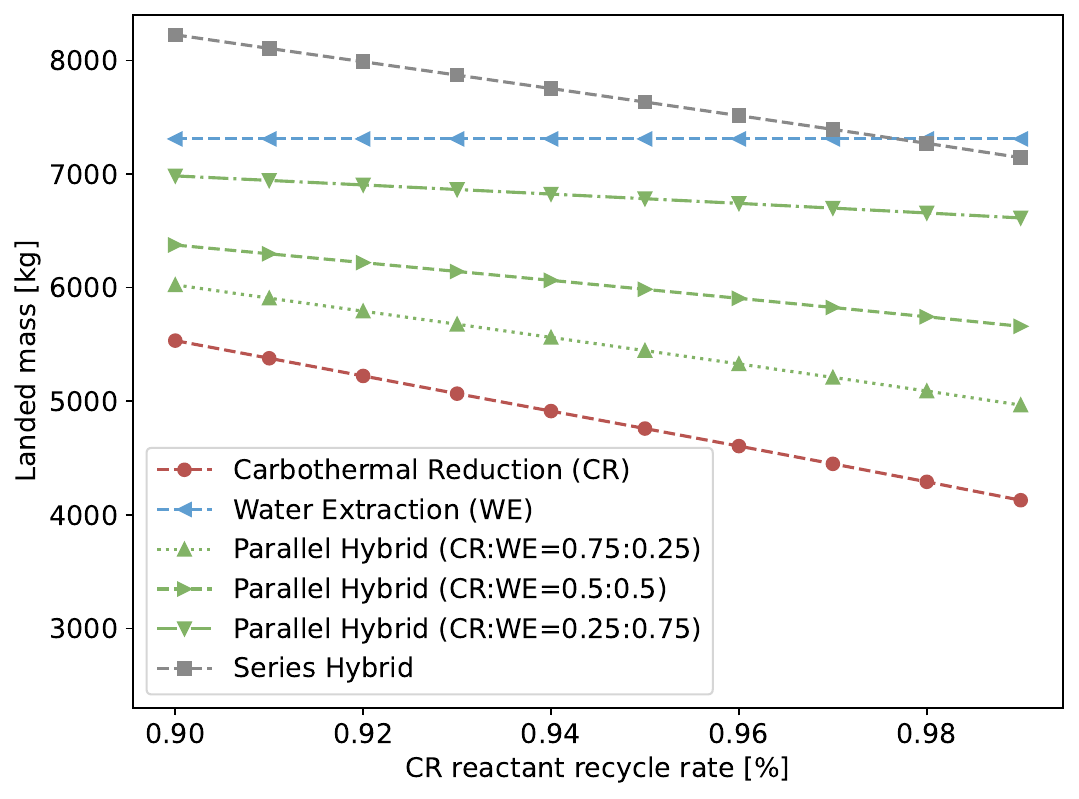}
         \caption{One-year operation.}
         \label{fig:sizing_mass_recycle_rate_1year}
     \end{subfigure}\\
     \vspace{0.25 cm}
     \begin{subfigure}[]{0.99\columnwidth}
         \centering
         \includegraphics[width=0.95\linewidth]{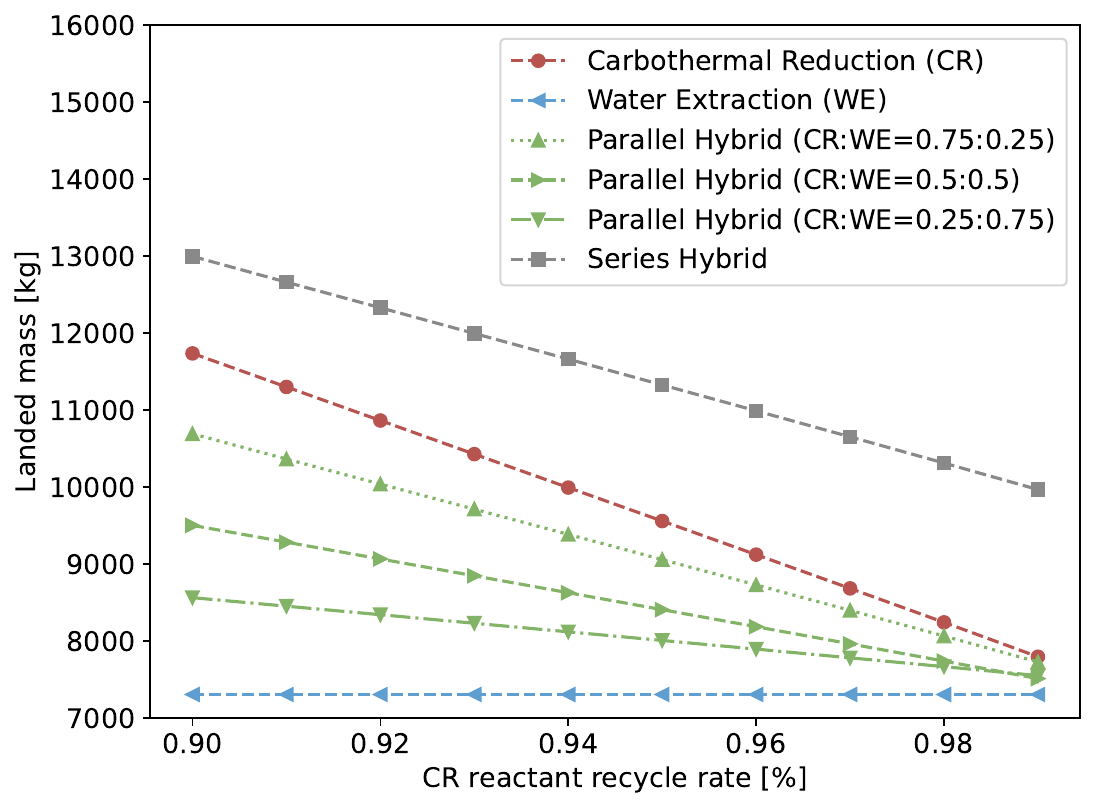}
         \caption{Three-years operation.}
         \label{fig:sizing_mass_recycle_rate_3years}
     \end{subfigure}
\caption{System mass as a function of the CR reactant recycle rate. The target production mass is set to 10.0 t of LOX and 1.25 t of LH\textsubscript{2}.}
\label{fig:sizing_mass_recycle_rate}
\end{figure}

\rev{Figure \ref{fig:sizing_mass_recycle_rate} shows the system mass dependence on the CR reactant, i.e., CH\textsubscript{4} and H\textsubscript{2}, recycle rate.
As can be seen in this figure, there is a linear relationship between the recycle rate and the landed mass.
For one year of operation (Fig.~\ref{fig:sizing_mass_recycle_rate_1year}), the CR can be the lightest architecture. For a higher recycling rate, the difference between  the WE and the \revaa{SH} architecture can be almost negligible. When the operational period is three years (Fig.~\ref{fig:sizing_mass_recycle_rate_3years}), the WE can be the lightest regardless of the CR reactant recycle rate.}

\begin{figure}[pos=h]
\centering
     \begin{subfigure}[]{\columnwidth}
         \centering
         \includegraphics[width=0.95\linewidth]{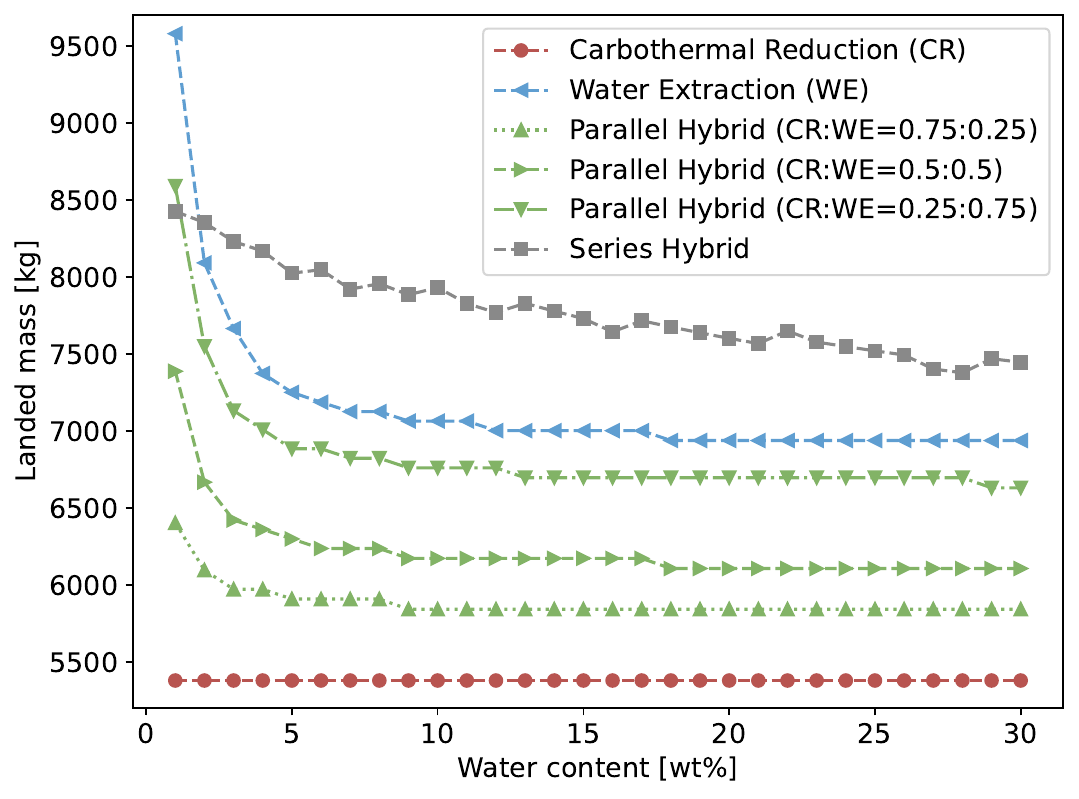}
         \caption{One year operation.}
         \label{fig:sizing_mass_water_content_1}
     \end{subfigure}\\
     \vspace{0.25 cm}
     \begin{subfigure}[]{\columnwidth}
         \centering
         \includegraphics[width=0.95\linewidth]{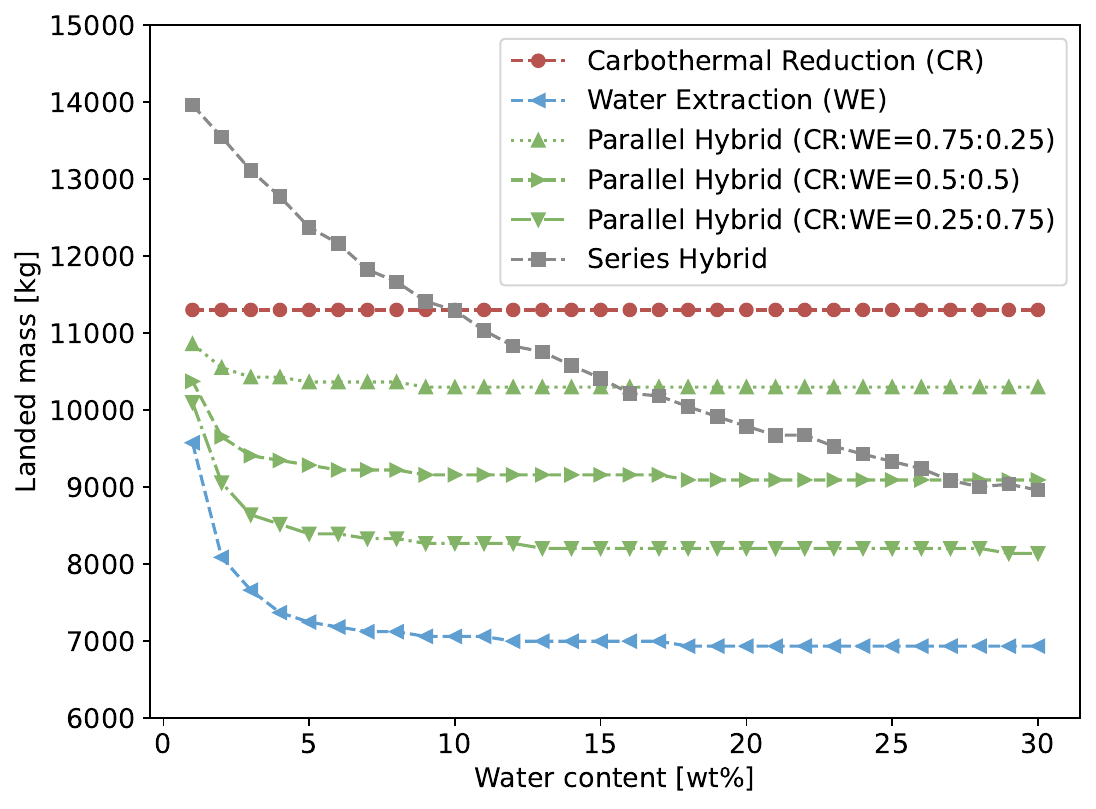}
         \caption{Three years operation.}
         \label{fig:sizing_mass_water_content_3}
     \end{subfigure}
\caption{System mass as a function of water ice content in regolith.}
\label{fig:sizing_mass_water_content}
\end{figure}

\begin{figure}[pos=h]
\centering
\includegraphics[width=0.95\linewidth]{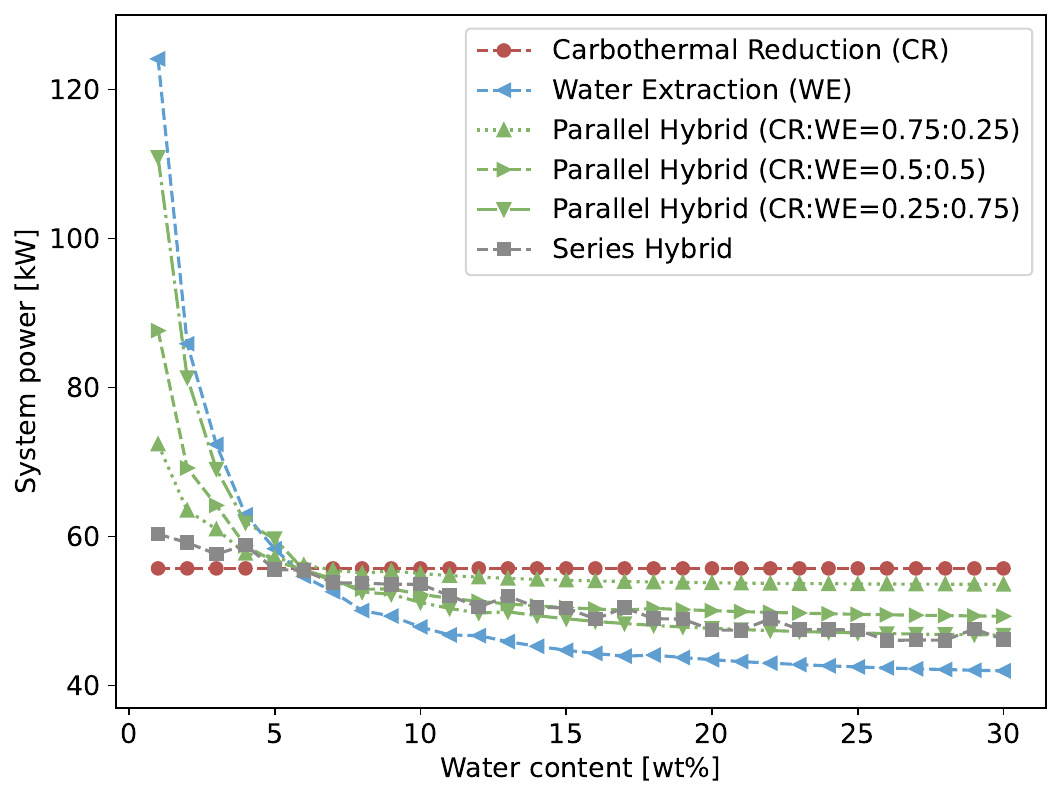}
\caption{\revaasecond{System power as a function of water ice content in regolith.}}
\label{fig:sizing_power_water_content}
\end{figure}

Water ice content in the excavated regolith in a PSR also affects the comparable landed mass and the power consumption of the WE and the hybrid plants significantly (Figs.~\ref{fig:sizing_mass_water_content} and \ref{fig:sizing_power_water_content}). Generally, as water content increases, both the landed mass and power consumption decreases. This trend affects the \revaasecond{heaviest} plant architecture, \revaasecond{as well as} the \revaasecond{most and the} least power-consuming plant architectures.
As can be seen in Fig.~\ref{fig:sizing_mass_water_content_1}, the water content has the largest impact on the mass of the WE architecture for one year operation. However, for three years' operation, the landed mass of the \revaa{SH} architecture becomes significantly lighter as the water content becomes larger.
This is because as the water content increases, the \revaa{SH} architecture depends more on the WE from the icy regolith rather than the CR of the dry regolith, making the system mass for dry regolith processing smaller. \rev{Since the CR architecture can be a lot heavier than the WE architecture for the three years of operation, the \revaa{SH} architecture can be lighter by depending less on the CR architecture.}

\revaasecond{Regarding power requirements, the WE architecture is affected the most by the water content due to its sole reliance on the WE process. As a result, the WE architecture consumes the most power when the water content is smaller than or equal to 4 wt\% while it becomes the least power-consuming architecture when the water content is 6 wt\% or higher. In contrast, the SH architecture requires relatively low power across the analyzed range. This is because the SH architecture significantly depends on the CR process more when the water content is low, and gradually shifts its reliance to the WE process as water content increases.}
The water content in the PSR is further considered for uncertainty analyses (Section~\ref{sec:water_content}).

\section{Uncertainty in lunar ISRU} \label{sec:uncertainty_in_lunar_isru}
\subsection{Uncertain parameters} \label{sec:uncertain_parameters}

\revaasecond{As indicated by the study by Cilliers et al.~\cite{Cilliers.2020}, there exists the risk of underperformance when various uncertainties, such as in resource content, are not considered.}
To assess the benefits and drawbacks of each ISRU plant architecture \revb{compared in Section \ref{sec:result_deterministic}}, this paper further analyzes the effects of uncertainties on the performance of the plant. \revaasecond{Inspired by the aforementioned study by Cilliers et al., }Monte Carlo simulations of 5000 scenarios \rev{with different values for parameters related to the lunar environment and ISRU operations} \revb{is utilized to }reveal the performance of each ISRU architecture \rev{under various conditions. }
Table \ref{tab:monte_carlo} summarizes the Monte Carlo parameters assumed in this study.

\begin{table}[t]
   \small
   \centering
   \caption{Monte Carlo parameters for uncertainty consideration.}
   \vspace*{3mm}
   \label{tab:monte_carlo} 
   \begin{tabularx}{\linewidth}{>{\raggedright}X  >{\raggedright\arraybackslash}X}
   \hline    \hline  
   \textbf{Variable} & \textbf{Distribution $(\mu, \sigma^2)$} \\ \hline 
   Particle size mass fraction {[}wt\%{]}       & $\mathcal{N}$(62.6, 8.3\textsuperscript{2})\\
   Silica content {[}wt\%{]}                  & $\mathcal{N}$(45.2, 0.93\textsuperscript{2})\\
   Water ice content {[}wt\%{]}                 & Nominal: \textit{Lognormal}\,(ln(4), 0.3\textsuperscript{2}) \\
                                            & Best: $\mathcal{N}$(30, 2\textsuperscript{2})\\
   Water yield {[}-{]}                 & $\mathcal{N}$(0.75, 0.05\textsuperscript{2}) \\
   Operational availability {[}days/year{]}           & $\mathcal{N}$(200, 10\textsuperscript{2})\\
   \hline  \hline  
   \end{tabularx}
\end{table}

\subsubsection{Particle size distribution of lunar highlands regolith}
\begin{figure}[bth!]
\centering
\includegraphics[width=.95\linewidth]{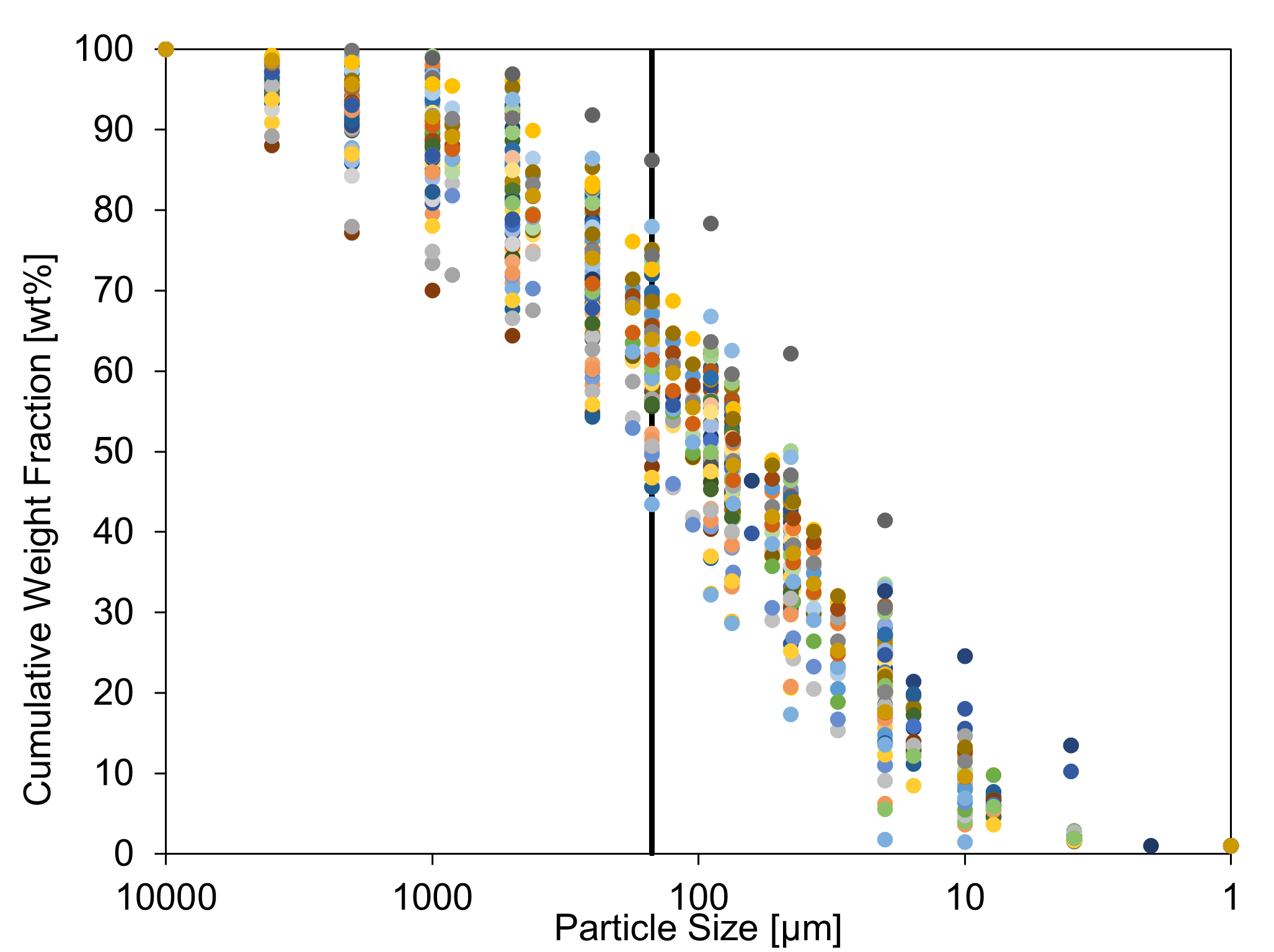}
\caption{Particle size distributions of 75 Apollo 16 samples obtained from \revaasecond{Graf}~\cite{Graf.1993}. The black vertical line represents 150 \textmu{}m.}
\label{fig:psd}
\end{figure}

\begin{figure}[bth!]
\centering
\includegraphics[width=.98\linewidth]{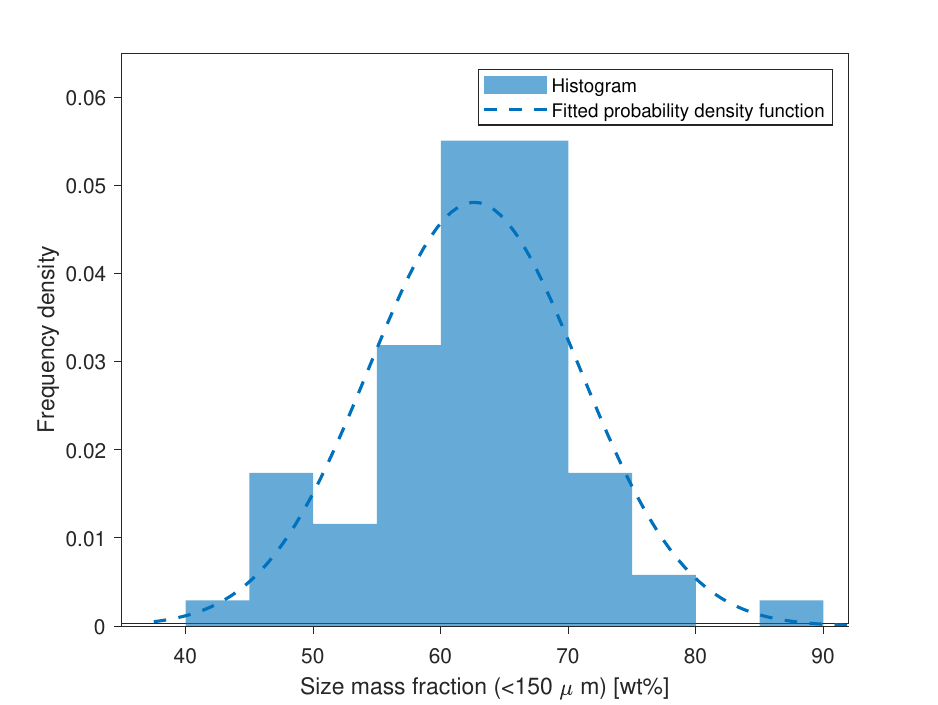}
\caption{Distribution of mass fraction of 75 Apollo 16 samples with a particle size smaller than 150 \textmu{}m.}
\label{fig:pdf_psd}
\end{figure}

After excavation, the particle size separator removes the coarse regolith fraction (Fig.~\ref{fig:architecture}). In this work, to improve the conversion time of the \revaa{CR}, it is assumed that particles larger than particles larger than 150 \textmu{}m are removed as mentioned in Sec.~\ref{sec:beneficiation}. A further assumption is that the lunar surface in the southern polar region is mainly covered by highlands regolith similar to Apollo 16 samples \citep{Korotev.2003, Cannon.2023}. Figure \ref{fig:psd} shows the particle size distributions of 75 Apollo 16 samples reported by \revaasecond{Graf}~\cite{Graf.1993}. From these distributions, Fig.~\ref{fig:pdf_psd} is generated to depict the distribution of the mass fraction of the excavated regolith between 90 and 1000 \textmu{}m. As can be observed in Fig.~\ref{fig:pdf_psd}, the mass fraction is normally distributed with a mean value of 62.6 wt\% and a standard deviation of 8.3 wt\%, i.e., $\mathcal{N}$(62.6, 8.3\textsuperscript{2}). Therefore, the mass fraction of excavated regolith in the southern polar region is assumed to follow this distribution.


\subsubsection{Silica content in lunar highlands regolith}\label{sec:silica_content_in_lunar_highlands_regolith}

As reported \revaasecond{by Ridley et al.~}\citep{Ridley.1973}, silica is the most dominant metal oxide in lunar highlands regolith. The Apollo 16 sample 60501 contains 45.24 wt\% of silica with a standard deviation of 0.93 wt\%. Without further information on the silica content of the lunar southern polar regolith, we also assumed $\mathcal{N}$(45.2, 0.93\textsuperscript{2}) as a distribution of the silica content.

\subsubsection{Water ice content in the lunar south polar region} \label{sec:water_content} 
Compared to the silica concentration, information regarding water ice content in the southern polar region of the Moon is significantly more limited. A great number of past studies have tried to estimate the icy water content in PSRs. For instance, Colaprete et al.~\cite{Colaprete.2010} analyzed Lunar Crater Observation and Sensing Satellite's ejecta plume and concluded that the water ice content in the impact site (Cabeus crater) was 5.6 $\pm$2.9 wt\%. The authors also noted that icy water might exist heterogeneously. Hayne et al.~\cite{Hayne.2015} supported this heterogeneity showing some cold traps, where water ice can be thermally stable, do not have exposed water ice. This also aligns with the finding of only 3.5\% of PSRs probably having surface ice by Li et al.~\cite{Li.2018}.
From UV albedo values, Hayne et al. concluded that 0.1–2.0 wt\% of water ice might be exposed in some PSRs if the ice is intimately mixed with dry regolith, whereas 1-10\% of surface area would be icy if this is pure and patchy.

More recently, using the data from the Moon Mineralogy Mapper, Li et al.~\cite{Li.2018} estimated the water content in PSRs by detecting near-infrared absorption features of water ice in reflectance spectra. This study revealed that ice-bearing pixels (280 m $\times$ 280 m) may contain around 30 wt\% of ice intimately mixed with regolith or 20 vol\% if pure and patchy ice exists. 
This is a good agreement with Zuber et al.~\cite{Zuber.2012}, who concluded that 20\% superficial thin ($\sim$1 \textmu{}m) ice may exist in the Shackleton crater. As Cannon and Britt~\cite{Cannon.2020} argued, estimations made by Hayne et al.~\cite{Hayne.2015} and Li et al. correspond to only surface ice (micrometers to millimeters depth), whereas the value measured from the Lunar Crater Observation and Sensing Satellite impact might be subsurface data (tenths of centimeters depth).

\begin{figure}[bt!] 
\begin{center}
\includegraphics[width=\linewidth]{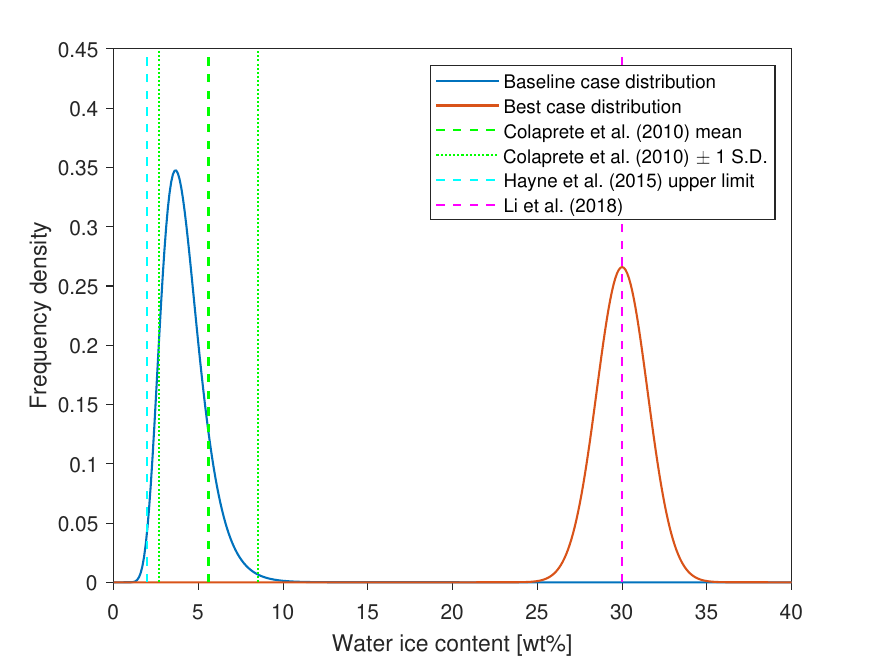}
\end{center}
\caption{Water ice content. \revaasecond{Values from past studies \cite{Colaprete.2010, Hayne.2015, Li.2018} and the distributions considered in this paper.}}\label{fig:pdf_ice}
\end{figure}

These past studies indicate significant heterogeneity of water ice in the southern polar region might be expected. 
With all the information available to this date, it is hard to conclude the water content in excavated icy-regolith. Therefore, this work tests two different scenarios: a nominal and a best-case scenario. In the nominal scenario, the large ice content reported by \revaasecond{Li et al.~}\cite{Li.2018} exists only on the surface, and the actual ice content in the excavated regolith is smaller than that, which aligns with Cannon and Britt~\cite{Cannon.2020}. For this scenario, to avoid negative values, a log-normal distribution with a median value of 4 wt\% and a variance of 0.3 wt\%\textsuperscript{2} is considered
as the distribution of the ice content. We also test the other scenario where there indeed is more ice in the excavated regolith as reported by \revaasecond{Li et al.} within a reachable area from the PEL site. For this scenario, $\mathcal{N}$(30, 2\textsuperscript{2}) [wt\%] is employed. See Fig. \ref{fig:pdf_ice} for employed distributions as well as some reported values in past studies.

\subsubsection{Water yield} \label{sec:water_yield}
The water yield efficiency of the \revaa{WE} technology is another uncertain parameter.
Kleinhenz and Paz assumed 75\% efficiency \cite{Kleinhenz.2020}.
Cole et al. reported 67 $\pm$ 5\% of \revaa{WE} in 25 minutes of heating from highland regolith simulant \cite{Cole.2023}. From these studies, this paper assumed 75\% mean efficiency with 5\% of variance.

\subsubsection{Operational availability} \label{sec:operational_availability}
Although this work does not prescribe a detailed power system, operational uncertainty related to the power source should be considered. Assuming solar power, operational availability depends on how much sunlight the plant can receive. Mazarico et al.~\cite{Mazarico.2011} simulated the sunlight availability of many PELs in both northern and southern polar regions (see supplementary tables 2a and 2b). According to their simulation, the best location near the ridge of the Shackleton crater can get about 202 days of continuous sunlight which aligns with the assumption made by Linne et al.~\cite{Linne.2021} Therefore, a normal distribution $\mathcal{N}$(200, 8\textsuperscript{2}) [days per year] is assumed in this paper.
\subsection{Lunar ISRU performance under uncertainty} \label{sec:lunar_isru_performance_under_uncertainty}

\begin{figure}[bth!]
\centering
\begin{minipage}{.48\textwidth}
\centering
\includegraphics[width=0.95\linewidth]{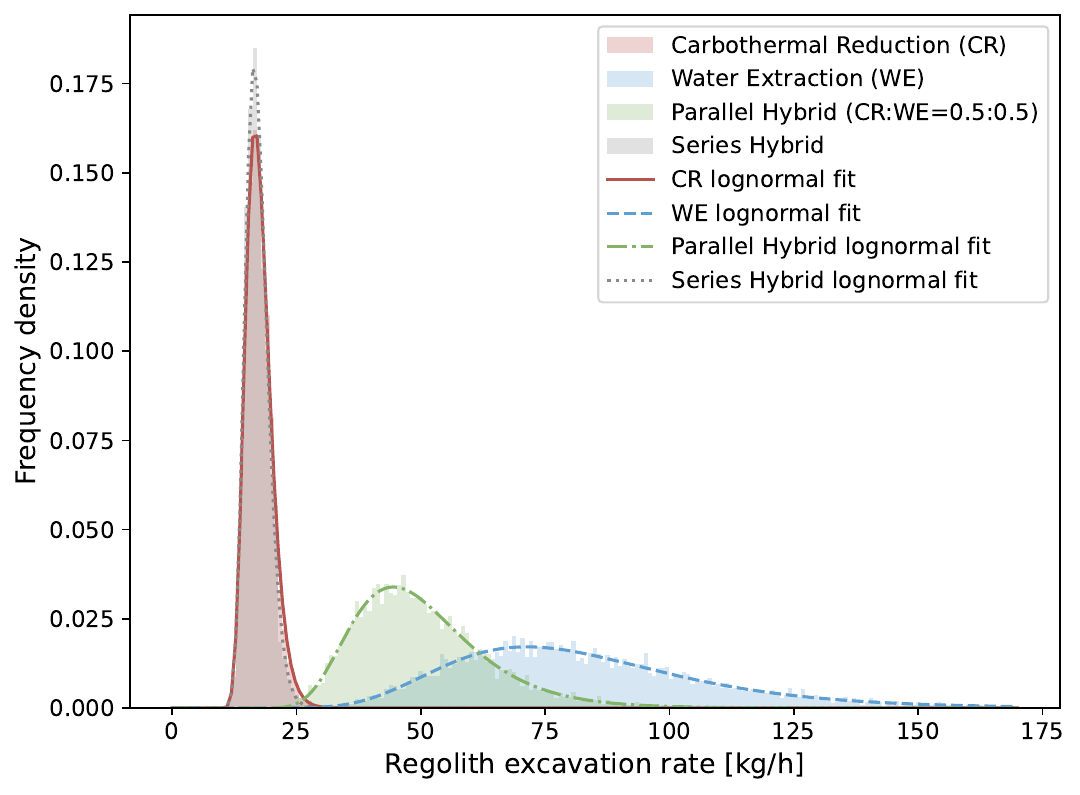}
\caption{Excavation rate to meet target production, 10 t of LOX and 1.25 t of LH\textsubscript{2}. \revaasecond{The nominal case for water ice content distribution: $\textit{Log-normal}\,(ln(4), 0.3\textsuperscript{2})$ [wt\%]}.}
\label{fig:mc_exc_rate}
\end{minipage}\hfill
\begin{minipage}{.48\textwidth}
\centering
\includegraphics[width=0.95\linewidth]{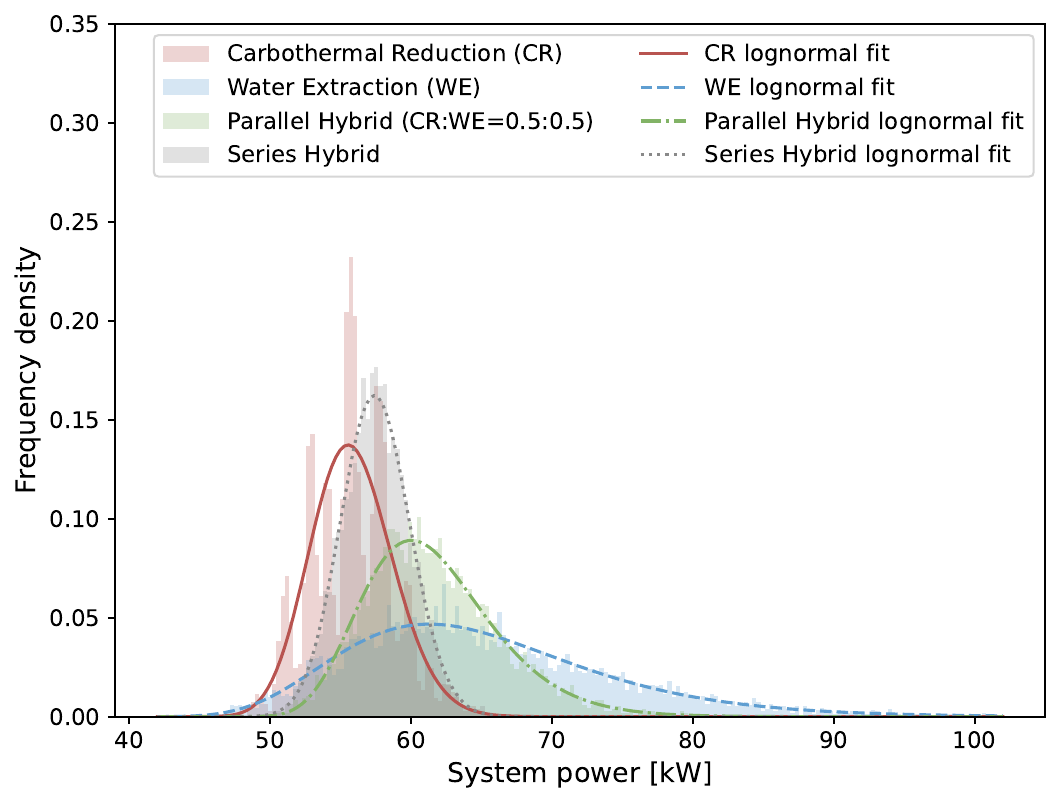}
\caption{Power consumption to produce 10 t of LOX and 1.25 t of LH\textsubscript{2}. \revaasecond{The nominal case for water ice content distribution: $\textit{Log-normal}\,(ln(4), 0.3\textsuperscript{2})$ [wt\%]}.}
\label{fig:mc_power}
\end{minipage}\hfill
\end{figure}

Using the varying parameters listed in Table~\ref{tab:monte_carlo}, we run Monte Carlo simulations for 5000 scenarios. \revaa{Each scenario selects a different parameter set following distributions listed in Table \ref{tab:monte_carlo}.}
Figure \ref{fig:mc_exc_rate} depicts the regolith excavation rate of each architecture. As can be seen in this figure, the CR and the \revaa{SH} architectures require much smaller excavation rate than the WE and the \revaa{PH} architectures. This is due to the high oxygen yield of the CR technology compared to the WE technology. The \revaa{SH} architecture can potentially achieve a \revaa{even} higher yield than the CR architecture by extracting oxygen from both regolith and water ice.
However, due to the distance between a PEL and PSR, hydrogen generated in the PEL is not recycled to \revaasecond{the methanation process (Equation \ref{eq:methanation})} in this architecture. This limitation leads to the yield of the \revaa{SH and CR} architecture \revaa{almost the same}.
The uncertainty in the water ice content makes the required regolith excavation rates for the WE and the \revaa{PH} architectures unclear. For instance, for the WE architecture, \revaasecond{79} kg/h of regolith excavation is required to produce 10 t of liquid oxygen and 1.25 t of liquid hydrogen with a 50\% confidence. However, to achieve a 95\% confidence, the required excavation rate becomes \revaasecond{132} kg/h. \rev{According to} \revaasecond{Mueller et al.~}\cite{Mueller.2016}, RASSOR is designed to excavate at least 2.7 t of regolith per day (112.5 kg/h). Therefore, depending on the condition, the WE architecture has a risk of not meeting the excavation mass target unless a redundant excavator is considered.

Figure \ref{fig:mc_power} shows the histograms of \revaasecond{the} power consumption of each architecture with varying parameters. As can be seen, the average required power of the \revaasecond{WE architecture is the highest and that of the CR is the lowest. The average power values of the SH and the PH are between those of the CR and WE}. Although the average power consumption of the WE architecture is \revaasecond{higher} than the other architectures, due to the larger uncertainty in the water ice content, the histogram of the WE architecture shows the widest variety in its power consumption. \revaasecond{The lower bound of power consumption for both the CR and WE architectures is similar. Specifically, at the 5\% confidence level, the required power for the WE architecture is approximately 52 kW—only about 1 kW higher than that of the CR architecture.} The non-normality observed in this figure is due to the non-continuity of some design parameters. For instance, the number of molten zones for the CR process and the number of cryocoolers to liquefy oxygen and hydrogen must be integers, making several peaks in the histograms.

\begin{figure}[bth]
\centering
\begin{minipage}{.48\textwidth}
\centering
\includegraphics[width=0.99\linewidth]{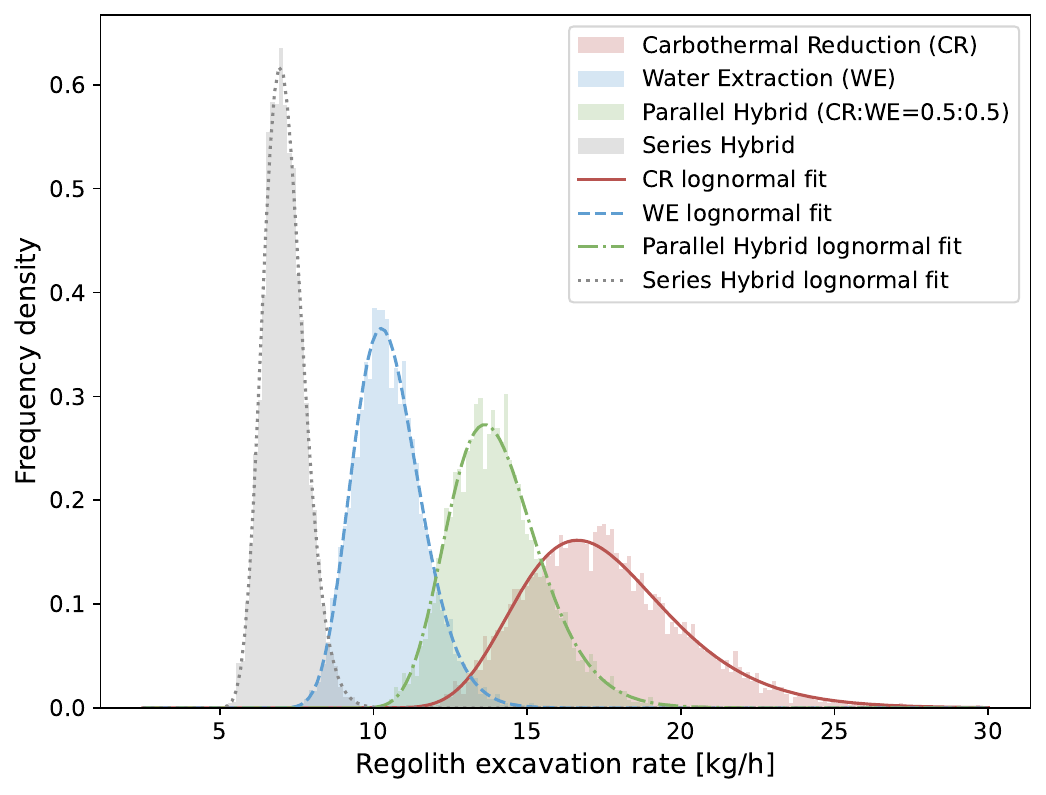}
\caption{Excavation rate to meet target production, 10 t of LOX and 1.25 t of LH\textsubscript{2}. The best case for water ice content distribution: $\mathcal{N}$(30, 2\textsuperscript{2}) \revaasecond{[wt\%]}.}
\label{fig:mc_exc_rate_30}
\end{minipage}\hfill
\begin{minipage}{.48\textwidth}
\centering
\includegraphics[width=0.99\linewidth]{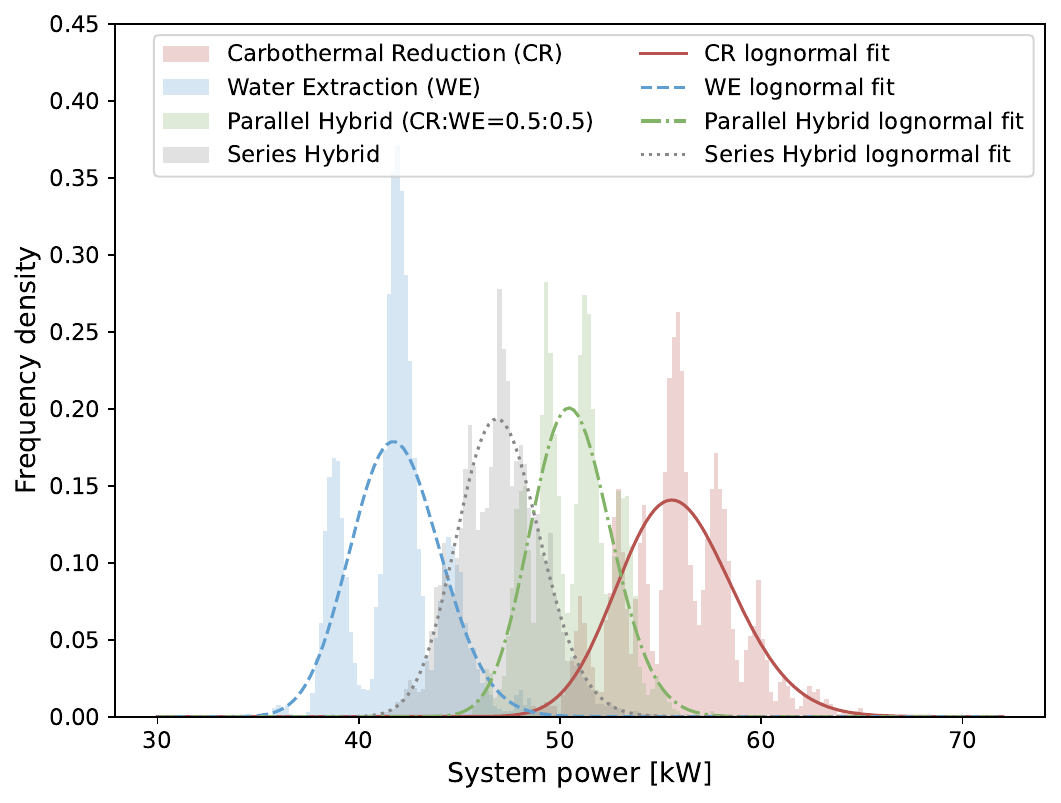}
\caption{Power consumption to produce 10 t of LOX and 1.25 t of LH\textsubscript{2}. The best case for water ice content distribution: $\mathcal{N}$(30, 2\textsuperscript{2}) \revaasecond{[wt\%]}.}
\label{fig:mc_power_30}
\end{minipage}
\end{figure}

Figures \ref{fig:mc_exc_rate_30} and \ref{fig:mc_power_30} are the results of the Monte Carlo simulations with the best-case water content (i.e., $\mathcal{N}$(30, 2\textsuperscript{2}) \revaasecond{[wt\%]}). With this abundance of water ice in the lunar southern pole, the the \revaa{SH} architecture \revaa{significantly} outperforms the other architectures in terms of the regolith excavation rate (Fig.~\ref{fig:mc_exc_rate_30}). For power consumption (Fig.~\ref{fig:mc_power_30}), the WE plant is expected to require the least \revaa{followed by the SH}. The similar non-normality seen in Fig.~\ref{fig:mc_power} is observable in Fig.~\ref{fig:mc_power_30}, too.

\begin{figure}[bth!]
\centering
     \begin{subfigure}[b]{0.48\textwidth}
         \centering
         \includegraphics[width=0.99\linewidth]{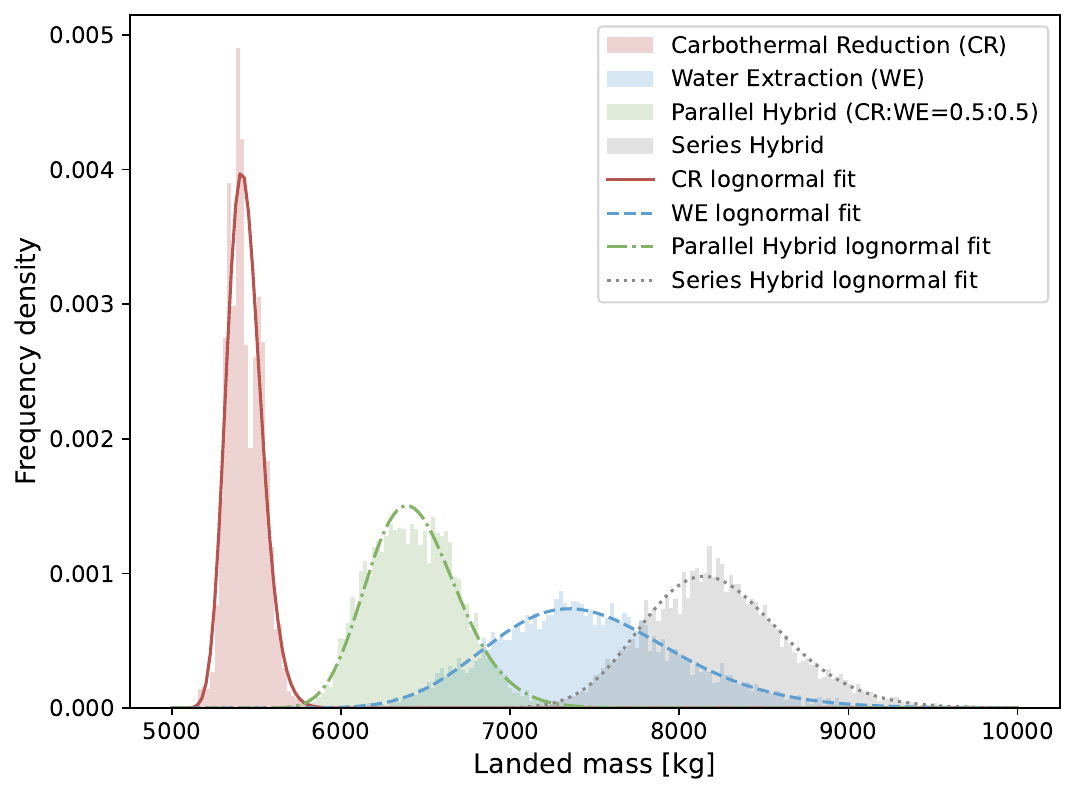}
         \caption{Water ice content: \textit{Log-normal}\,(ln(4), 0.3\textsuperscript{2}) \revaasecond{[wt\%]}.}
         \label{fig:mc_mass_4}
     \end{subfigure}\hfill
     \begin{subfigure}[b]{0.48\textwidth}
         \centering
         \includegraphics[width=0.99\linewidth]{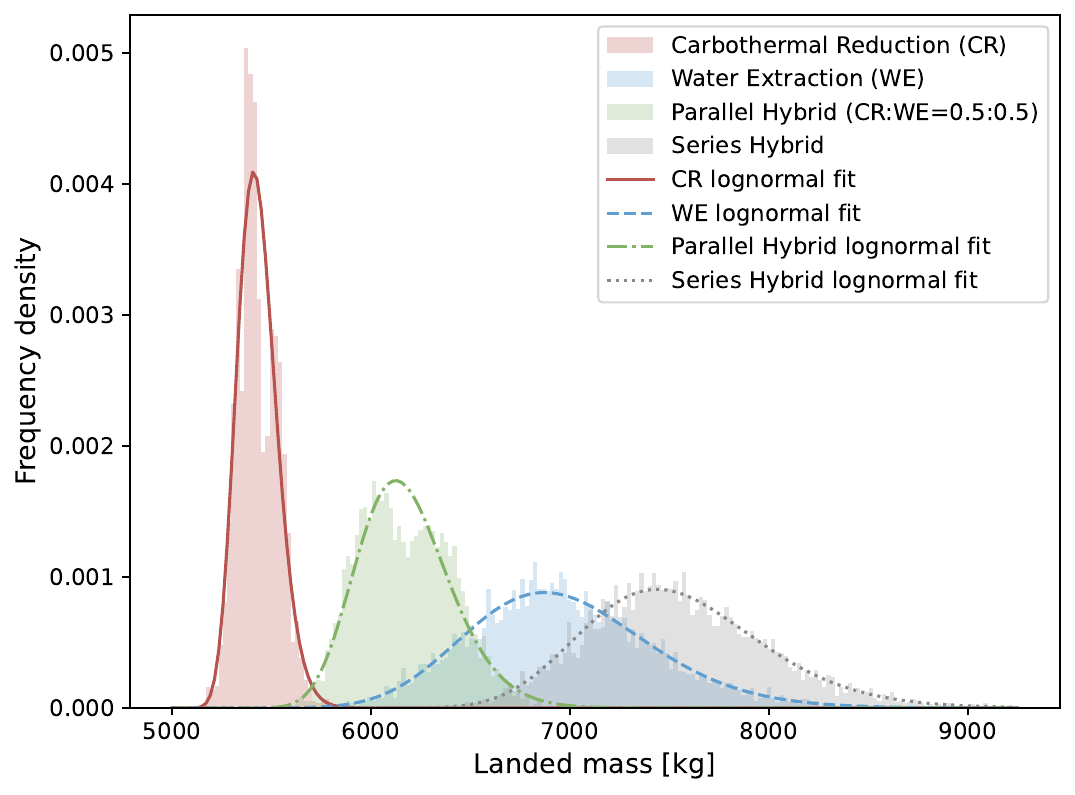}
         \caption{Water ice content: $\mathcal{N}$(30, 2\textsuperscript{2}) \revaasecond{[wt\%]}.}
         \label{fig:mc_mass_30}
     \end{subfigure}
\caption{Mass requirement to meet target production of 10 t of LOX and 1.25 t of LH\textsubscript{2}.}
\label{fig:mc_mass}
\end{figure}

Finally, the required comparable mass including the additional hydrogen and methane for one year's operation are summarized in Fig.~\ref{fig:mc_mass}. For both water ice content distributions, the CR plant mass variance is significantly smaller than the others. In terms of the expected system mass, the lightest architecture is the CR while the heaviest is the \revaa{SH} architecture. 

\section{Discussion} \label{sec:discussion}
\subsection{Summary of key findings}
As can be seen in Fig.~\ref{fig:sizing_mass_1}, the CR, architecture appears to be the lightest for one year operation. This result is aligned with what Kleinhenz and Paz concluded~\cite{Kleinhenz.2020}, although the mass of additional methane mass and its storage due to the imperfect reactant recycling \rev{is often overlooked in the literature}.

For power consumption, \revaasecond{while the CR architecture requires the least power in the baseline case (Fig.~\ref{fig:sizing_power}), it is revealed that the promising architectures change drastically depending on the water content (Figs.~\ref{fig:sizing_power_water_content}, \ref{fig:mc_power}, and \ref{fig:mc_power_30}).}
It should be noted that the CR technology in this paper requires direct thermal power instead of electrical power. \rev{This can be particularly challenging for the \revaa{SH} architecture since the CR process of this hybrid architecture is conducted in a PSR. This requires a mirror system suggested \revaasecond{by Sowers and Dreyer} \cite{Sowers.2019}, indicating more landed mass is required compared to the other architectures.}

From the Monte Carlo studies (Sec.~\ref{sec:uncertainty_in_lunar_isru}), the WE architecture turns out to require the largest excavation rate range, highlighting the challenge of engineers to optimize its design. This can also be seen in Fig.~\ref{fig:mc_mass} as the widest range of the required system mass for the WE plant.
Table~\ref{tab:mass_95} summarizes the required mass with a 50\% confidence to meet the target production rate and those with a 95\% confidence.
To improve the robustness of each plant design, 95\% confidence is more desirable. \revaa{With the nominal assumption of water content, the WE architecture requires} \revaasecond{1,017} kg \revaa{of additional mass to achieve this confidence while the CR architecture only requires about} \revaasecond{185 kg.}

\begin{table}[bth!]
\caption{Baseline mass and additional mass for the improved robustness.}
\label{tab:mass_95}
\begin{subtable}{\columnwidth}
\caption{Water ice content: \textit{Log-normal}\,(ln(4), 0.3\textsuperscript{2}) \revaasecond{[wt\%]}.}
\centering
\resizebox{\columnwidth}{!}{%
\begin{tabular}{lrrrr}
\hline\hline
Confidence level to meet  & \multicolumn{4}{c}{Comparable mass {[}kg{]}}\\
the target production & CR & WE &  PH&  SH\\
\hline
50\% & \revaasecond{5,414} & \revaasecond{7,428} &\revaasecond{6,429}  &\revaasecond{8,202}  \\
95\%  &\revaasecond{5,600} & \revaasecond{8,445} &\revaasecond{6,897}  &\revaasecond{8,956}\\
\hline
\hline
\end{tabular}}
\end{subtable}\\
\vspace{0.25 cm} 
\begin{subtable}{\columnwidth}
\caption{Water ice content: $\mathcal{N}$(30, 2\textsuperscript{2}) \revaasecond{[wt\%]}.}
\centering
\resizebox{\columnwidth}{!}{%
\begin{tabular}{lrrrr}
\hline\hline
Confidence level to meet & \multicolumn{4}{c}{Comparable mass {[}kg{]}}\\
the target production & CR & WE &  PH&  SH\\
\hline
50\% & \revaasecond{5,415} & \revaasecond{6,932} &\revaasecond{6,159}  &\revaasecond{7,506}  \\
95\%  &\revaasecond{5,596} & \revaasecond{7,785} &\revaasecond{6,574}  &\revaasecond{8,347}\\
\hline
\hline
\end{tabular}}
\end{subtable}
\end{table}

\subsection{Benefits and risks of each plant architecture}
\subsubsection{Carbothermal reduction architecture}
The CR plant \revaa{requires a slow excavation rate (Fig.~\ref{fig:mc_exc_rate}) due to the high oxygen yield of the process.}
Moreover, the CR plant can be the lightest architecture when the operational period is one year (Figs.~\ref{fig:sizing_mass_target_ox_1}, \ref{fig:sizing_mass_recycle_rate_1year}, and \ref{fig:sizing_mass_water_content_1}) \revaasecond{and the least power-consuming architecture in comparison depending on the water content (see Figs.~\ref{fig:sizing_power_water_content} and \ref{fig:mc_power})}. However, this architecture depends the most on the import of resources from Earth (Fig.~\ref{fig:sizing_mass}), which can make it significantly heavier (and costly) when the operational period is longer (Fig.\ref{fig:sizing_mass_prod_period}).

\subsubsection{Water extraction architecture}
As the Monte Carlo analyses reveal, the WE plant has the widest range in the required regolith excavation rate (Fig.~\ref{fig:mc_exc_rate})\revaasecond{, required power (Fig.~\ref{fig:mc_power}), and landed mass (Fig.~\ref{fig:mc_mass}) While it can still be the lightest plant architecture for operations lasting two years or longer (Fig.~\ref{fig:sizing_mass_prod_period}), achieving greater system robustness requires significantly more mass (Table~\ref{tab:mass_95}).}

If the water content in the excavated regolith is indeed as high as 30 wt\% as reported \revaasecond{by Li et al.~}\cite{Li.2018}, the expected regolith excavation becomes smaller (Fig.~\ref{fig:mc_exc_rate_30}) and the expected power consumption becomes \revaa{the smallest} (Fig.~\ref{fig:mc_power_30}). \revaasecond{However, if the water content is low, this architecture demands substantially higher power (Fig.~\ref{fig:sizing_power_water_content}). Therefore, decision-makers must carefully evaluate this option and ensure greater certainty in water content before implementation.}

\subsubsection{Parallel hybrid architecture}
The performance, indicated by the power consumption, and regolith excavation, of the \revaa{PH} architecture turned out to be between that of the CR and the WE architectures \revaasecond{(see e.g., Figs.~\ref{fig:sizing_power_target_ox} and \ref{fig:sizing_mass_water_content}).}
The required landed mass for the \revaa{PH} architecture is not the lightest. However, the mass difference between the \revaa{PH} architecture and the CR or the WE architecture \rev{not always significant} (see e.g., Fig.~\ref{fig:sizing_mass_target_ox}), indicating that one can gather a lot more information that cannot be obtained from the CR or WE architecture by adding a little mass. Therefore, having a hybrid pilot plant to gather more information to help decision-makers make more informed decisions for a full-scale plant could be beneficial. This potential flexibility in the operation and deployment of a hybrid plant should be explored in the future.

\subsubsection{Series hybrid architecture}
The obvious benefit of the \revaa{SH} architecture is the small regolith excavation rate with a small variance regardless of the water ice content (Figs.~\ref{fig:mc_exc_rate} and \ref{fig:mc_exc_rate_30}).
This merit comes with a cost as the required mass which is often the heaviest among the examined architectures \revaasecond{(see e.g., Fig.~\ref{fig:mc_mass})}. This mass cost can become even larger when we consider the technological challenge of performing the CR process in a PSR. 

\revaasecond{Regarding  power consumption, this architecture is expected to require relatively small power in comparison regardless of the water ice content (Figs.~\ref{fig:sizing_power_water_content}, \ref{fig:mc_power}, and \ref{fig:mc_power_30}). Therefore, for risk-averse decision-makers, this architecture would be advisable.}

\subsection{\revaa{Limitations and direction for future work}}

\revaa{In this paper, the performances of different lunar ISRU architectures are compared quantitatively. As Kiewiet et al.~\cite{Kiewiet.2022} discussed, however, qualitative aspects of each technology, such as technological maturity should also be considered as part of the criteria.
The maturity of the technologies and the complexity of each design were considered as criteria \revaasecond{by Taylor and Carrier} \cite{Taylor.1992b}. Similar analysis to these studies should be conducted in the future to help decision-makers consider other perspectives.}

\revaa{Furthermore, there are risks inherent in each plant architecture. All considered architectures employ water electrolysis to produce oxygen and hydrogen. There exists a risk of potential suspension of the products when a technological malfunction occurs. The benefit of deploying multiple smaller-scale plants similar to \revaasecond{the study by Linne et al.}~\cite{Linne.2021} should be explored as well.}

\revaa{For the SH architecture, rejecting a large heat load in a PSR can affect the environment in multiple ways. It could lead to the sublimation of icy water in the PSR. It also could affect the icy regolith underneath the processing plant. Without carefully assessing these risks, the SH architecture should not be deployed on the Moon.}

\revaasecond{Global optimizations of the entire design and operations are left for future work. Adjusting certain parameters, such as the size of the molten zone in the CR process, may lead to a more optimal design. Additionally, design parameters can affect the operations of each architecture, potentially impacting both average and peak power consumption.}

\revaa{This paper also does not explore the flexibility in the operation enabled by a hybrid plant. Deploying a hybrid plant can gather more data than a single technology plant, indicating the better decisions to be made next. This flexible operation will be discussed in the future.}

\revaasecond{For a more detailed ISRU system design, improving the fidelity of certain subsystem models is desirable. Incorporating a parametric sizing model for the methanation chamber and integrating various beneficiation technologies may provide decision-makers with additional insights into system design and operations.}

\revaa{Finally, the inclusion of power systems and consideration of different technologies, such as microwave heating of icy regolith, would help decision-makers to better understand the trade space between different architectures. A similar approach to this paper, subsystem-level modeling, and uncertainty analysis, can be taken for such an extension, as well.}


\section{Conclusion} \label{sec:conclusion}

This paper has proposed a new lunar ISRU production plant architecture integrating both carbothermal reduction of dry regolith and water extraction from icy regolith. 
Two different types of hybrid plants are explored. The parallel hybrid architecture has two mining sites in both a PEL and a PSR processing regolith simultaneously. In the series hybrid architecture, dry regolith tailing from the water extraction is further processed by carbothermal reduction.
The potential concept of operations of the proposed hybrid architectures is discussed in detail.

The proposed hybrid plant architectures are compared with conventional carbothermal reduction and water extraction plants in terms of mass and power. The total mass and power are estimated from the subsystem-level models of each plant design.
Some key parameters that affect the mass and power significantly, such as water ice content, are detected and investigated. Furthermore, Monte Carlo simulations of 5000 scenarios have detected the range of required excavation rate and the power consumption of each plant design.

The uncertainty in water ice content largely affects the required excavation rate \revaasecond{as well as power consumption}. This suggests the urgent need to gather more information on lunar water ice. The series hybrid architecture turned out to be a reasonable choice in terms of the regolith excavation rate showing its merit regardless of the water ice content. \revaasecond{This architecture also consumes relatively low power among all considered architectures.} However, the series hybrid architecture becomes the heaviest architecture in many cases.

Although this paper focuses on the plant designs, the potential flexible operations enabled by the proposed hybrid design should be explored further. By gathering more information on both technologies and their operations in the lunar environment, a hybrid pilot plant may provide crucial information for engineers and mission architects. It can help, ultimately, inform better decisions for full-scale plant design and deployment in the coming decades.

\section*{Acknowledgement}
\revaa{The authors would like to thank Joshua N. Rasera, Ph.D., Luka Malone, and} \revaasecond{three} \revaa{anonymous reviewers for their careful reading and thoughtful suggestions for improvement.}

\printcredits

\setcounter{table}{0}
\renewcommand{\thetable}{A.\arabic{table}}
\onecolumn
\appendix
\section{Comparisons with existing mass and power estimation}\label{sec:mass_power_comparison}

This \rev{appendix} compares the mass and power of each subsystem estimated from the proposed model with values reported in \revaasecond{two past studies}~\cite{Linne.2021, Kleinhenz.2020}. Table \ref{tab:cr_comparison} lists the mass and power of each subsystem estimated using the proposed model and reported in Linne et al.~\cite{Linne.2021}. Linne et al. scaled their plant using three independent plants with each producing 3.5 t of oxygen. To compare with this model, we scaled a plant to produce 3.5 t of oxygen and multiplied the mass and power by three.
\revaa{There are large differences in material transporting mass. In this paper, we have only considered an auger-based particle size separator, while multiple longer regolith transportation systems are considered by Linne et al. The PEM stack, too, shows a large difference in mass between our estimation and \revaasecond{theirs}. Note that the model used by Linne et al. is unclear, and thus, further assessment of the PEM stack is required.}

\begin{table}[pos = b]
\small
\caption{Comparison of the Carbothermal Reduction Plant with Linne et al.~\cite{Linne.2021}}
\label{tab:cr_comparison}
\centering
\begin{tabular}{lrrrr}
\hline\hline
 & \multicolumn{2}{c}{Mass {[}kg{]}${}^{\ast 1}$} &  \multicolumn{2}{c}{Power {[}W{]}${}^{\ast 1}$}  \\
Element & Est. value  &  Linne et al.~\cite{Linne.2021} & Est. value & Linne et al.~\cite{Linne.2021} \\ \hline
CR reactor &\revaa{572}&  177&  \revaasecond{48,270}& 43,290 \\
Desulfurization &  \revaa{(not modeled)}&78&\revaa{(not modeled)}& 20\\
Methanation reactor & 51& 42& -&-\\
Recycle loop compressor &\revaa{1}&2& 204& 1,482\\
Heat exchanger &\revaa{5}&17&-& -\\
Condenser/electrolyzer H\textsubscript{2}O tank &\revaa{(not modeled)}&28&-& -\\
System gas tank w/ compressor &\revaa{(not modeled)}&18& \revaa{(not modeled)}&  450\\
H\textsubscript{2} tank & \revaa{10}& 8& -&- \\
CH\textsubscript{4} tank &\revaa{3}&8& -&-\\
Hoppers &\revaa{94}&123& -&-\\
Materials size-sorting and transporting &\revaa{2}&227& \revaa{12}&\revaa{610}\\
Electrolysis pump &\revaa{3}&6& \revaa{128}&\revaa{450}\\
PEM stuck &\revaa{8}&\revaa{152}& 16,145&9,150\\
Dryers &\revaa{1}&\revaa{4}& -&-\\
Electrolysis valves and lines &\revaa{(not modeled)}&38& -&-\\
Excavator &223  &  \revaa{276}&\revaasecond{1,266}&-\\\hline\hline
 \multicolumn{5}{r} {$\ast 1 $: All values include 30\% growth~\cite{AIAA.2015}.}
\end{tabular}
\end{table}

Table \ref{tab:we_comparison} lists the mass and power of each subsystem estimated using the proposed model and reported in \revaasecond{Kleinhenz and Paz}~\cite{Kleinhenz.2020}. To compare these values, the assumptions here are adopted from \revaasecond{this reference. For instance, } the water content used here is 5 wt\%, and the operational availability is adjusted to 225 days per lunar year. The ratio of O\textsubscript{2} and H\textsubscript{2} here is 6:1; therefore, the plant is scaled to produce 13.3 t of LOX and 1.7 t of LH\textsubscript{2}.
The mass difference in the water extractor is due to the water extraction method; our extractor model is based on a batch operation, whereas Kleinhenz and Paz assumed continuous water extraction. 
The required number of excavators is one based on our model, which is also different from the two excavators' operation \revaa{proposed by Kleinhenz and Paz}.
\revb{The mass of gas dryers is modeled based on the in-house developed hardware at Johnson Space Center in \revaasecond{their study}, which causes large mass difference from the model developed in this paper.}
There is another large difference in the total mass of water tankers due to the number of tankers. While  Kleinhenz and Paz assumed only two tankers are required, our model outputs a lighter total mass when 10 tankers are used. This mass can be further lighter by increasing the maximum number of tankers; however, operating a large number of tankers could increase the complexity of the total system. 

\begin{table}[bth!]
\caption{Comparison of the direct water extraction plant with Kleinhenz and Paz~\cite{Kleinhenz.2020}.}
\label{tab:we_comparison}
\small
\centering
\begin{tabular}{lrrrr}
\hline\hline
 & \multicolumn{2}{c}{Mass {[}kg{]}${}^{\ast 1}$} &  \multicolumn{2}{c}{Power {[}W{]}${}^{\ast 1}$}  \\
Element & Est. value  & Kleinhenz and Paz~\cite{Kleinhenz.2020} & Est. value & Kleinhenz and Paz~\cite{Kleinhenz.2020} \\ \hline
Water extractor &\revaa{487}& 335& \revaa{25,014}& \revaa{20,749}\\
Excavator & 86& 160& \revaa{487}& \revaa{211}\\
O\textsubscript{2} liquefaction &  \revaa{159}& 153& \revaa{4,800}&\revaa{4,093}\\
O\textsubscript{2} cryostorage &  \revaa{727}& 213& -& -\\
H\textsubscript{2} liquefaction & \revaa{1923}& 1780& \revaa{25,515}& \revaa{21,171}\\
H\textsubscript{2} cryostorage & \revaa{902}& 299& -& -\\
H\textsubscript{2} \& O\textsubscript{2} dryers & 1& 52& -& \revaa{58}\\
PEM electrolysis & \revaa{7}& 48& \revaa{22,192}& \revaa{22,042}\\
Water tank & \revaa{266}& 123&  -& \revaa{199}\\
Water tankers & 3742 & 1868& \revaa{2,633}& PEL: \revaa{375}\\
 &  & & & PSR: \revaa{534} \\ \hline\hline
\multicolumn{5}{r} {$\ast 1 $: All values do not include 30\% growth to compare them with Kleinhenz and Paz~\cite{Kleinhenz.2020}.}
\end{tabular}
\end{table}
\twocolumn


\bibliographystyle{model1-num-names}
\bibliography{references}






\end{document}